\documentclass{egpubl}
\usepackage{url}
\usepackage{fancyvrb}
\usepackage{fvextra}
\usepackage{graphicx}

\usepackage[T1]{fontenc}
\usepackage{dfadobe}  

\usepackage{cite}  
\usepackage{xcolor}
\BibtexOrBiblatex
\electronicVersion
\PrintedOrElectronic
\graphicspath{{figs/}}
\usepackage{egweblnk}
\usepackage{subcaption}
\usepackage{algorithm}
\usepackage{algpseudocode}
\usepackage{authblk}

\title[AVA]{AVA: Towards Autonomous Visualization Agents through Visual Perception-Driven Decision-Making}

\makeatletter
\renewcommand*{\@fnsymbol}[1]{\ensuremath{\ifcase#1\or *\or \dagger\or \ddagger\or
   \mathsection\or \mathparagraph\or \|\or **\or \dagger\dagger
   \or \ddagger\ddagger \else\@ctrerr\fi}}
\makeatother

\author[S. Liu,  H. Miao, Z. Li, M. Olson, V. Pascucci \& P-T. Bremer]
{
  \parbox{\textwidth}{\centering Shusen Liu$^{1}$\thanks{Contribute equally}\thanks{Project lead}, Haichao Miao$^{1*}$, Zhimin Li$^{2}$, Matthew Olson$^{1}$, Valerio Pascucci$^{2}$ and Peer-Timo Bremer$^{1,2}$}

  \parbox{\textwidth}{\centering $^1$Lawrence Livermore National Laboratory, $^2$SCI Institute, University of Utah}
}

\begin{document}

\teaser{
 \includegraphics[width=0.9\linewidth]{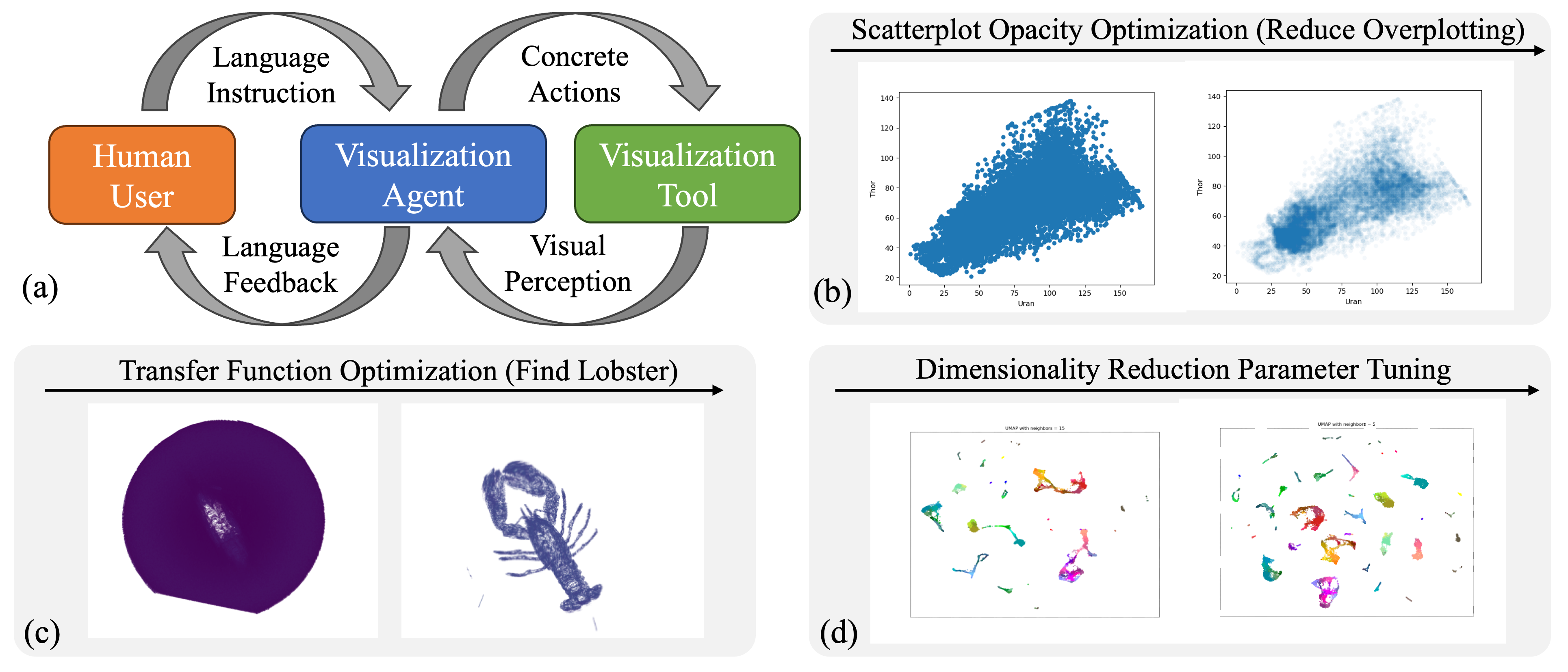}
 \centering
  \caption{Overview of Autonomous Visualization Agents (AVAs): By combining natural language understanding with visual perception (a) AVAs can not only understand user instructions but also control and adjust a visualization system by interpreting its visual outputs to accomplish user-specified goals. We demonstrate the broad applicability of the proposed paradigm in multiple distinct scenarios including scatterplot opacity selection (b), volume rendering (c), and hyperparameter tuning for nonlinear dimensionality reduction (d).}
\label{fig:teaser}
}

\maketitle

\begin{abstract}
With recent advances in multi-modal foundation models, the previously text-only large language models (LLM) have evolved to incorporate visual input, opening up unprecedented opportunities for various applications in visualization.
Our work explores the utilization of the visual perception ability of multi-modal LLMs to develop Autonomous Visualization Agents (AVAs) that can interpret and accomplish user-defined visualization objectives through natural language. We propose the first framework for the design of AVAs and present several usage scenarios intended to demonstrate the general applicability of the proposed paradigm. 
The addition of visual perception allows AVAs to act as the virtual visualization assistant for domain experts who may lack the knowledge or expertise in fine-tuning visualization outputs. Our preliminary exploration and proof-of-concept agents suggest that this approach can be widely applicable whenever the choices of appropriate visualization parameters require the interpretation of previous visual output. 
Feedback from unstructured interviews with experts in AI research, medical visualization, and radiology has been incorporated, highlighting the practicality and potential of AVAs.
Our study indicates that AVAs represent a general paradigm for designing intelligent visualization systems that can achieve high-level visualization goals, which pave the way for developing expert-level visualization agents in the future.

\end{abstract}  



\section{Introduction}
Within just a few months, large language models (LLMs)  have been widely adapted to solve a variety of tasks~\cite{wang2023voyager, wu2023autogen, song2023llm, wang2023survey}. 
In the visualization domain, these models have been used to produce visualizations~\cite{dibia2019data2vis, dibia2023lida} either through visual grammars like \emph{Vegalit}~\cite{2017-vega-lite}, or directly generating visualization code (e.g., in Matplotlib \cite{hunter2007matplotlib}, VTK \cite{sullivan2019pyvista}). However, due to the inherently visual nature of these systems, purely language-based models have limited capability to make sense of their output. This significantly hampers or even prevents the analysis of the results and thus severely limits the opportunity for iterative interactions with the given visualization systems. The recent introduction of multimodal LLMs, such as GPT-4V, has the potential to address this fundamental limitation by filling the visual perception gap, which opens many possibilities for new paradigms of interaction between existing visualization tools and human users.

One particularly interesting and powerful usage is the adoption of an Autonomous Visualization Agent (AVA) that can act as the medium between domain experts and visualization tools to facilitate and enrich user interaction (see Figure \ref{fig:teaser}(a)). Here, the AVA is defined as an entity that can understand high-level instructions (i.e., natural language) and autonomously carry out a sequence of actions in a visualization system based on the model's prior knowledge. More specifically, given the ability to perceive the visualization output an AVA can adjust and refine the parameters to meet the initial user-specified goal.
Such an agent will not only be able to relieve the user from potentially tedious and repetitive tasks but will also accomplish non-trivial visualization goals by iterative refining the existing visualization through visual feedback (following the \textit{visualization -> perception -> action} paradigm, just as a visualization expert would do). 

The agent-assisted visualization paradigm has the potential to fundamentally change how users interact with existing and future visualization tools.
Despite enormous efforts from the visualization community to design user-friendly approaches, many standard visualization tools remain feature-rich and challenging to navigate for a wide range of users with diverse backgrounds. For example, many experts in the application domains (e.g. industrial or medical) continue finding that designing an effective transfer function \cite{ljung2016state} for volume rendering is a non-intuitive and challenging process \cite{fujishiro1999automating}. 
In other words, there often remains a fundamental knowledge gap between the developers who design the visualizations and the intended target users of such systems. Visualization agents tailored for each tool can act as virtual assistants that bridge this knowledge gap and enable non-visualization experts to easily control, steer, and iterate the visualization based on high-level objectives specified by natural language. 

\begin{figure} [!htbp]
    \centering
    \includegraphics[width=0.5\textwidth]{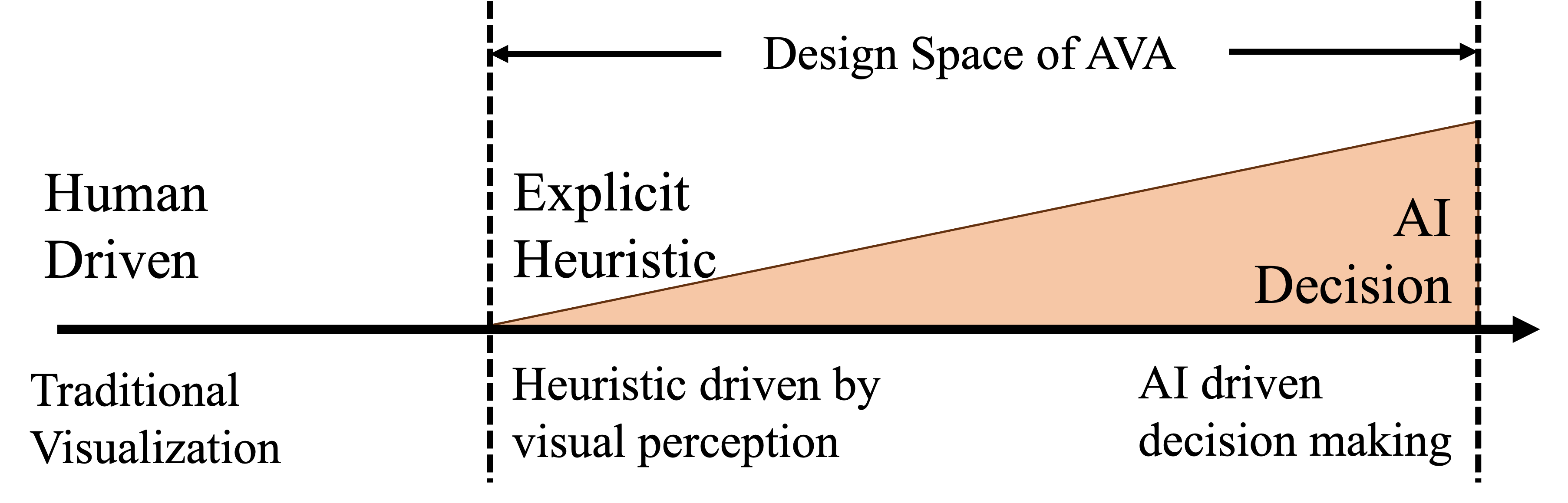}
    \caption{The design space of AVAs. 
    On one end, we explicitly encode heuristics on how to update the visualization parameter, i.e., how a transfer function should be changed, which is driven by a high-level objective specified through language, i.e., "does this show the structure of interest". Alternatively, we can aim for a fully self-directed system with no explicit guidance on its action beyond the initial instruction (prompt). }
    \label{fig:agent_design_space}
\end{figure}

Here, we aim to take the first step towards making AVAs a reality by exploring their design space and demonstrating their capability for solving real-world visualization tasks across different visualization areas.
The key power of AVAs derives from their ability to detect visual features associated with natural language instruction. Consequently, they can evaluate complex objectives that cannot be easily expressed algorithmically, i.e., \textit{is there a particular structure in the rendering results?} or \textit{does overplotting exist in the given scatterplot?}  
Despite its power, visual perception capabilities are only part of an agent. Once we obtain the visual understanding, the agent needs to plan its actions to achieve the goal. 
As illustrated in Figure \ref{fig:agent_design_space}, this presents a range of possible designs.
On one end (i.e., more explicit control), we can rely on heuristics to dictate the response. This is achieved by encoding our prior domain knowledge into decision rules. Alternatively, we can rely on the LLM and its prior knowledge to process the observations and plan the next action in a fully self-directed fashion.


To design an effective AVA, we first need to understand the capability and limitation of visual perception of the state-of-the-art multimodal LLMs (we use GPT4-Vision in all of our studies). We carried out a preliminary exploration of a few perception tasks related to common visualization outputs, including volume rendering, scatterplots, parallel coordinate plots, and graphs.
Leveraging what we learned from these simple benchmarks, we avoid areas of visualization where the visual perception of the current models is performing less accurately (e.g., graph, parallel coordinate).

Since different visualization tasks require potentially vastly different knowledge and strategies, instead of designing a general agent for arbitrary tasks, we approach the problem by designing specialized AVAs for different use cases under a common base agent implementation.
To demonstrate the feasibility and broad applicability of the proposed scheme, we intentionally select a distinctive set of applications, ranging from scientific/medical visualization to information visualization and dimensionality reduction. Our key contributions are:
\begin{itemize}
    \item Introduce AVAs, a new paradigm that leverages the visual perception capability of a machine learning model for autonomous decision-making in visualization. Make the first step toward building visualization agents that can act as virtual visualization experts;
    \item Provide a preliminary exploration of state-of-the-art multimodal LLM's visual perception ability for interpreting different visualization outputs, including, scatterplots, parallel coordinate plots, graphs, volume rendering outputs, etc.
    \item Demonstrate the feasibility and wide adaptability of AVA on several distinct visualization applications.
\end{itemize}






\section{Related Works}
\subsection{Visualization Generation and Recommendation}
Several existing tools explore how to generate visualizations based on user instructions. Data2vis~\cite{dibia2019data2vis} utilizes a recurrent network to generate code for visualization (e.g., with Matplotlib \cite{wang2023llm4vis}, or VTK \cite{sullivan2019pyvista}). The NL4DV~\cite{narechania2020nl4dv} approach turns visualization queries into visualization descriptors within the Vega-lite grammar \cite{2017-vega-lite}, and the work by Mitra et al.~\cite{mitra2022facilitating}, explores the back and forth interaction with such visualizations using a natural language interface. 
LLMs have been adopted for a similar role in LIDA~\cite{dibia2023lida}. 
The KG4Vis work~\cite{li2021kg4vis} adopts knowledge graphs to produce visualization recommendations. Besides just generating the visualization, several methods explore utilizing machine learning (ML) to help explain the rationale behind why a given visualization is recommended. For example, AdaVis~\cite{zhang2023adavis} adopts an attention-based model for explainable visualization recommendation, whereas the follow-up work leverages LLMs \cite{wang2023llm4vis} to achieve a similar goal.
With all such adaptions of LLMs in the visualization pipeline, it is also important to understand the width and depth of their capabilities beyond more constrained problems and contexts. Chen et al.~\cite{chen2023beyond} evaluate LLMs for solving visualization coursework by directly feeding them assignment descriptions.
Apart from generating code that produces visual output, LLMs are ideal for text description generation. Zong et al.~\cite{zong2022rich} utilize LLMs to generate descriptions of visualization for visually impaired users to understand and navigate the visualization.

Compared to these previous works, we are utilizing LLM in the proposed work in two significantly different ways. Firstly, our primary goal is to develop an agent that can refine visualization iteratively to accomplish specific visualization tasks, the objective of the agent is to understand and build upon an existing visualization, rather than focus on generating the visualization in the first place. In other words, we make the role of the agent as a visualization user rather than a visualization designer. Second, all existing works do not utilize the visualization output as input for the ML system for subsequent analysis, which significantly limits the capability and flexibility of the system. To the best of our knowledge, the proposed AVA is the first work that utilizes the visual perception of a multimodal LLM for visual analysis and autonomous decision-making.

\subsection{LLM-based Autonomous Agents}
With recent advances in LLM, there has been an explosive interest in developing LLM-based autonomous agents.
Compared to traditional reinforcement learning agents that often need to develop world understanding from scratch, LLM's in-depth prior knowledge and information processing capability make them more adaptive to complex environments and solving intricate tasks. 
Voyager \cite{wang2023voyager} introduces an LLM-powered embodied agent in Minecraft that can continuously explore the world and achieve milestones in the game world that were not possible with previous reinforcement learning approaches.
Due to the vast literature in this space and relevance to the current work, we refer readers to a comprehensive survey on LLM-based agents \cite{wang2023survey}.
In the following discussion, we will focus on vision task agents-related works.

Even though most LLM models are not designed from the ground up for processing visual inputs, many recent works try to incorporate external vision model \cite{suris2023vipergpt} or develop auxiliary components and fine-tune the model to provide additional vision capability \cite{feng2023layoutgpt, zhang2023gpt4roi}.
ViperGPT \cite{suris2023vipergpt} developed an agent that is capable of dividing the task into individual API calls to an external vision model for answering image-based queries and beyond. 
The layoutGPT \cite{feng2023layoutgpt} work introduced visual understanding for generating and reasoning about object placement in images and 3D scenes.
Gpt4roi \cite{zhang2023gpt4roi} augment LLM for fine-grained spatial reasoning capabilities.

These adaptions often focus on specific tasks and are trained on smaller-scale data, therefore are not designed for more general capabilities. 
This changes with the recent introduction of the GPT4-V (vision) \cite{openai2023gpt4vision} model by OpenAI, which added visual perception to one of the largest and most capable LLM models.
A detailed evaluation of a broad spectrum of visual understanding tasks is discussed in the ``The Dawn of LMM'' work \cite{yang2023dawn}.
Compared to the more general evaluation task in \cite{yang2023dawn}, we try to explore the GPT4-V model's visual perception capabilities on a specially designed set of visualization tasks, and eventually design an agent that is capable of refining and improving visualization output autonomously. 
\section{Background on Multi-modal Models and LLM Agents}

\noindent\textbf{Multi-Modal LLM.} Recently, we saw increasing popularity for models that can take multiple modalities as input \cite{radford2021learning} or models with input in one modality while output in another modality (e.g., text-to-image \cite{rombach2022high, ramesh2022hierarchical}, and image-to-text \cite{yu2022coca}).
In the context of this work, we focus on multi-modal LLMs \cite{yin2023survey} that can take both image and text as input.
LLMs, due to their capability and scale both in terms of parameters and training data size, are often referred to as foundation models.
These foundation models have access to a broad range of knowledge for understanding implicit context and common sense that was not possible before.
Since humans interact with the environment through multi-modal sensory input, the multi-modal LLM is an inevitable evolution of the text-only LLM systems. 
At the time of writing, the state-of-the-art multi-modal LLM is the GPT4-V (Vision) model, which the proposed work has utilized. However, it is important to note that the proposed AVA is not tied to a specified implementation of multi-modal LLM. 


\noindent\textbf{LLM Agents.}
In the context of machine learning, we can refer to an agent as a system or program that can autonomously make decisions or perform actions based on its environment.
An agent should be able to take action and make observations of its environment, and then reason about these observations before deciding on the subsequent action.
The action of the LLM agent can be in the form of textual or image output, however, their capabilities can be greatly enhanced by allowing them (e.g., ToolLlama \cite{qin2023toolllm} ) to make direct API function calls to utilize external tools.
When utilizing an LLM as the brain of the agent, we can enhance its action planning capabilities by adopting some simple protocols, such as chain-of-thoughts \cite{wei2022chain} (i.e., instruct it to solve the problem ``step-by-step''), or ReACT \cite{yao2022react} that connection the reasoning process with action the agent can take. 

\section{Preliminary Exploration of Multimodal-LLM for Static Visualization Perception}

\label{sec:vison_eval}
Before we can design an effective visualization agent that relies on visual input for decision-making, it is crucial to obtain some basic understanding regarding its capabilities and limitations for perceiving various types of visualization output.
It is important to note that this is \textbf{not intended to be a rigorous and systematic evaluation of multi-modal LLM ability}, as an in-depth study will require substantial resources and effort that is beyond the scope of this work. 
We hope this assessment can help illustrate what type of visualization agents we can realistically design and what would be the ideal tasks for such agents.

\noindent\textbf{Volume Rendering.}
We begin with evaluating the LMM's ability to recognize structures of interest within direct volume rendering images. 
Unlike photo-realistic images, that was explored by the work of Yang et al. \cite {yang2023dawn}, the outputs of volume rendering are subject to additional complexity (i.e., varying transparencies) introduced by the underlying transfer function.
To assess the model's capability, we present the model with the task of examining a screenshot and determining whether a specific object or structure of interest is 'recognizable'  or 'not recognizable'. 
We define the two assessments for the prompt as follows: \textbf{recognizable}: \textit{The structure of interest and its shape can be discerned in the screenshot.} \textbf{not recognizable}: \textit{The structure of interest cannot be identified in the image, even if another structure is recognizable.}

\begin{figure}[!htbp]
 \centering
 \begin{subfigure}{0.24\linewidth}
   \centering
   \includegraphics[width=\linewidth]{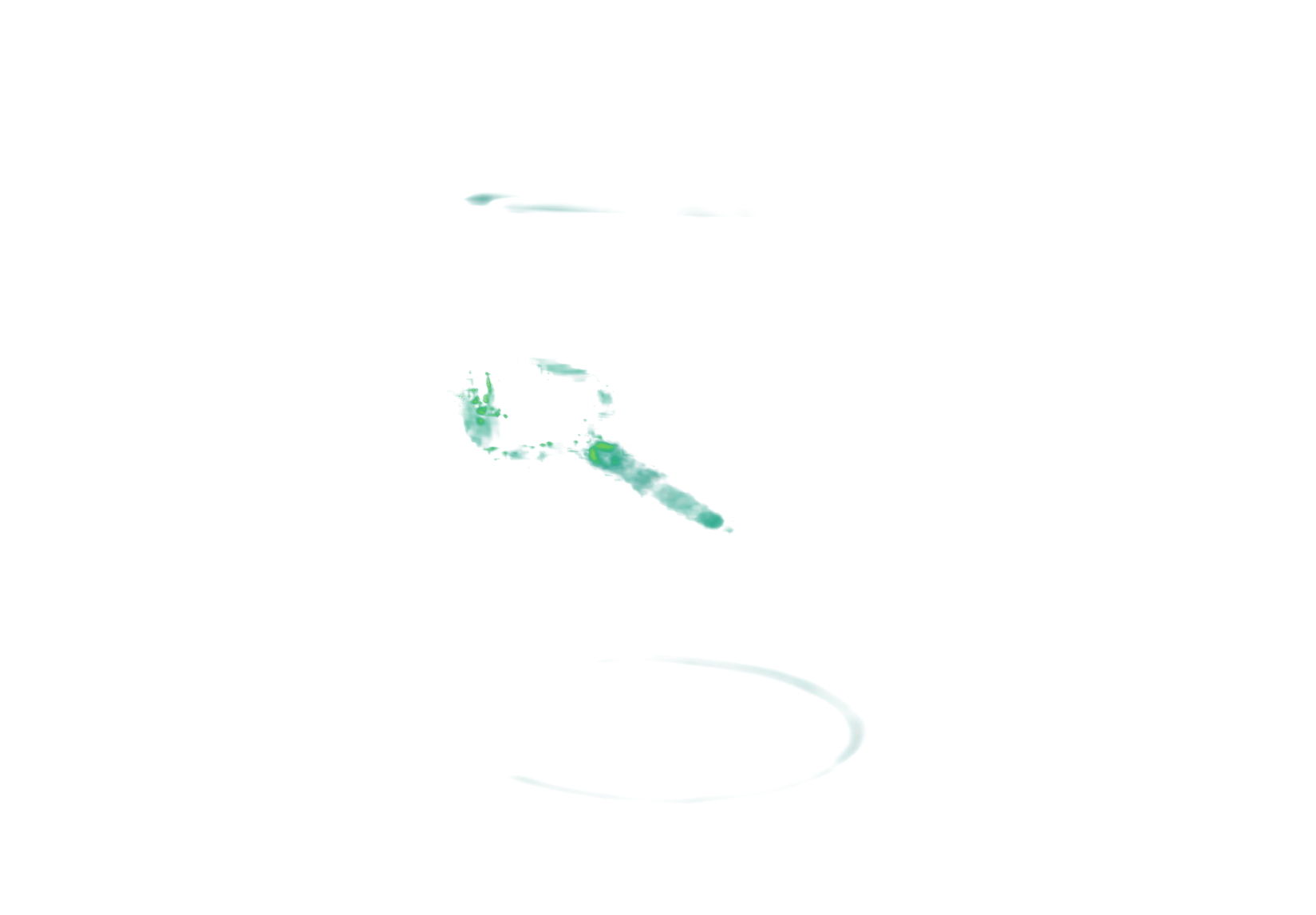}
   \caption{90\% }
   \label{fig:evalteapota}
 \end{subfigure}
 \begin{subfigure}{0.24\linewidth}
   \centering
   \includegraphics[width=\linewidth]{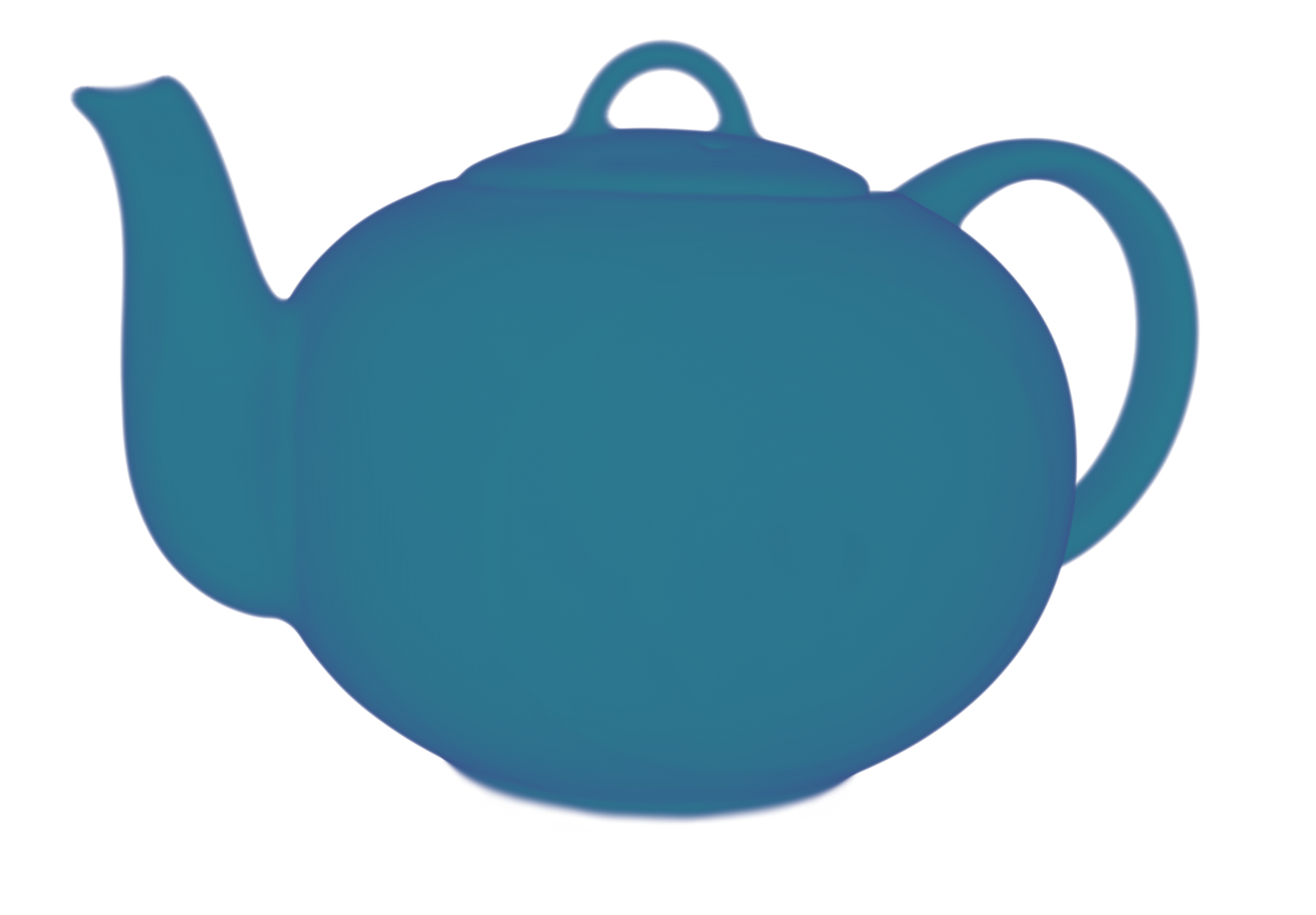}
   \caption{40\% }
   \label{fig:evalteapotb}
 \end{subfigure}
 \begin{subfigure}{0.24\linewidth}
   \centering
   \includegraphics[width=\linewidth]{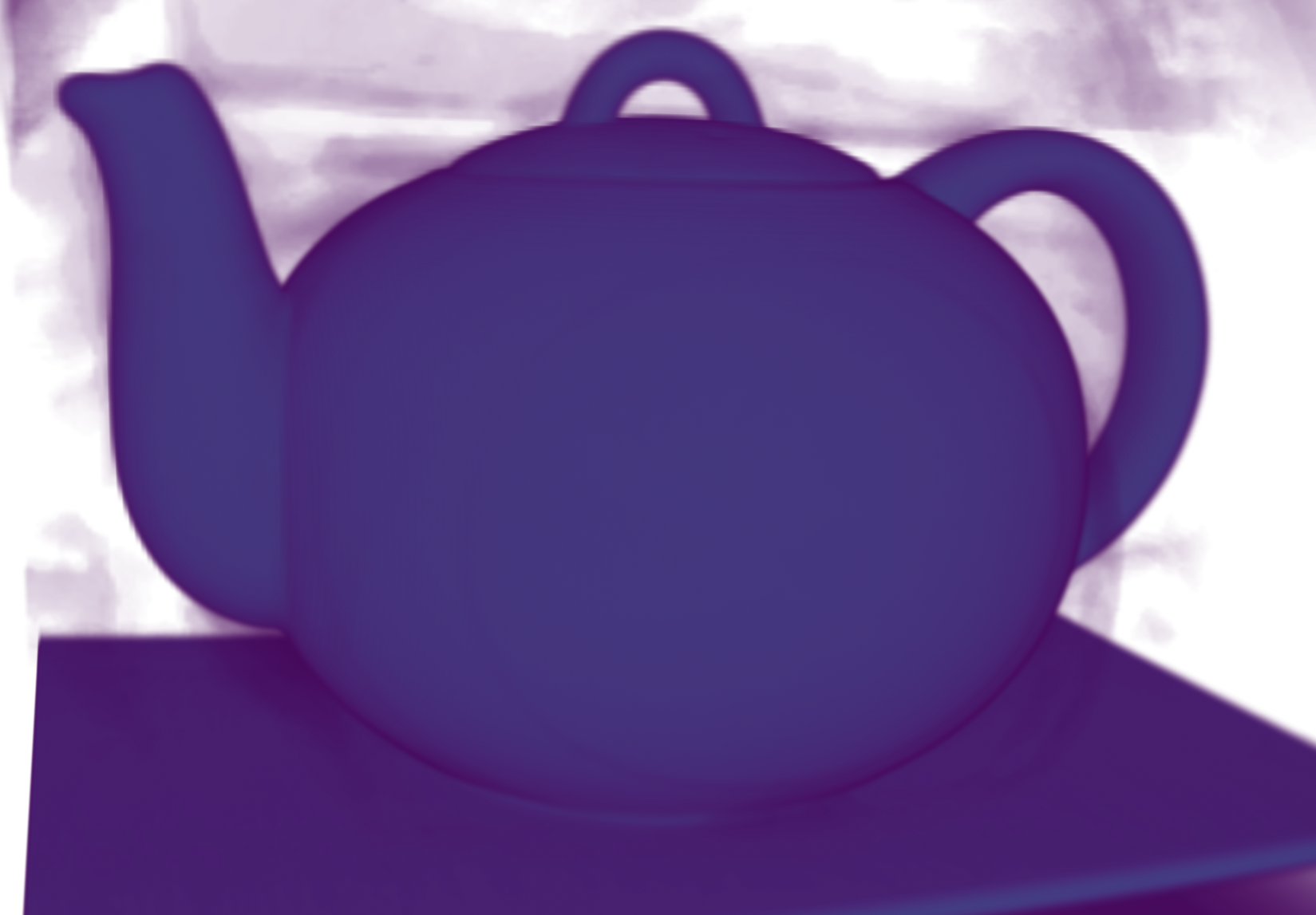}
   \caption{20\% }
   \label{fig:evalteapotc}
 \end{subfigure}
 \begin{subfigure}{0.24\linewidth}
   \centering
   \includegraphics[width=\linewidth]{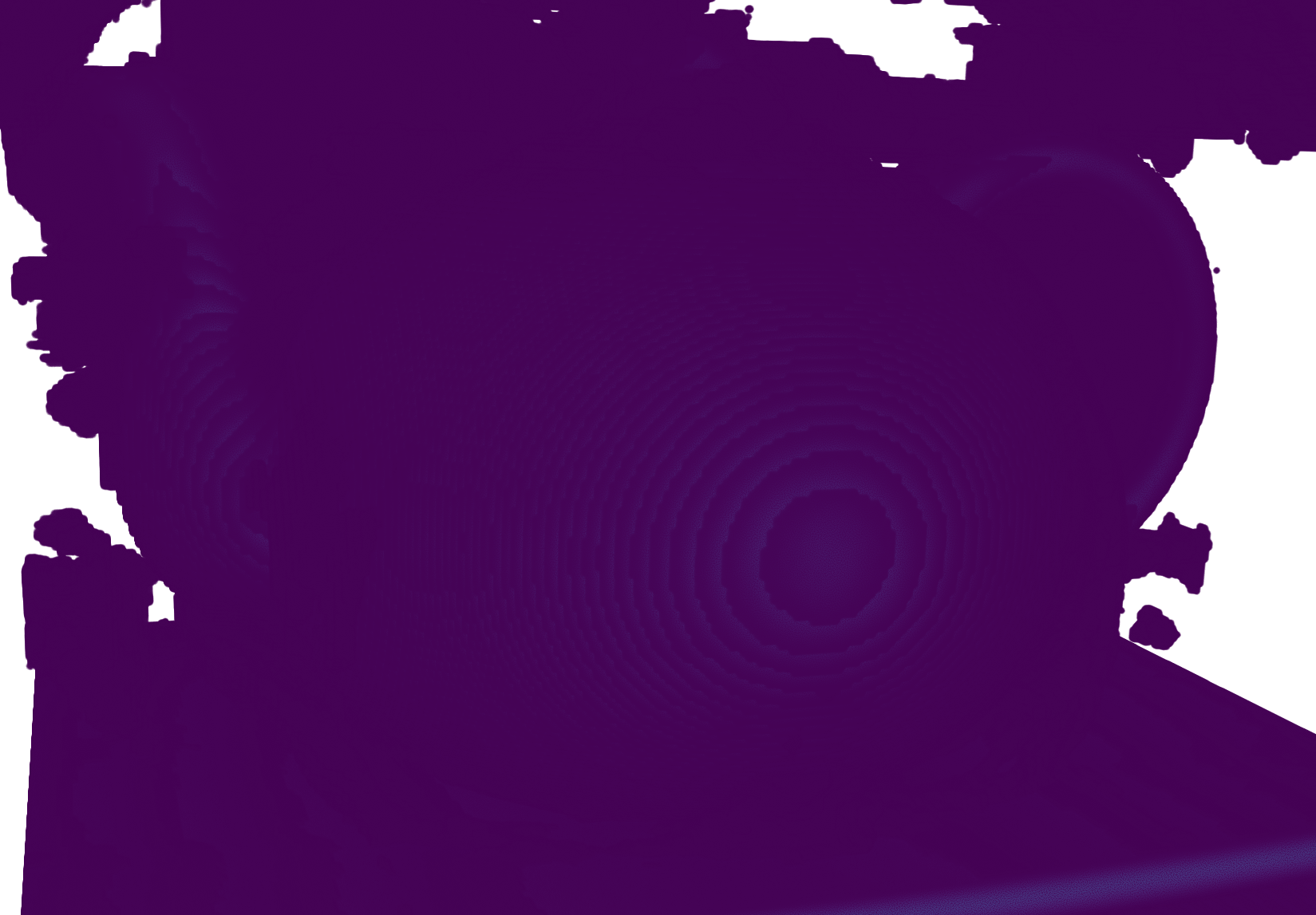}
   \caption{5\% }
   \label{fig:evalteapotd}
 \end{subfigure}
  
 \caption{The Boston Teapot dataset volume rendered using the same color map but at varying opacity levels. Structure of interest: the teapot. The response from the LLM model was \ref{fig:evalteapota}: 'not recognizable', \ref{fig:evalteapotb}: 'recognizable', \ref{fig:evalteapotc}: 'recognizable', and \ref{fig:evalteapotd}: 'not recognizable'}
 \label{fig:evalteapot}
\end{figure}

We assess the model on two datasets, the Boston Teapot \cite{teapot}
, and a downsampled version of the Visible Male \cite{spitzer1996visible}.
The reason why the Boston Teapot was selected for this experiment is because there is another structure, a lobster, located inside the teapot. 
As illustrated in Figures \ref{fig:evalteapot}, we maintained fixed rendering parameters, including viewpoint and colormap, while introducing variations solely in the opacity transfer function. To maintain uniformity across experiments, we employed a fixed-width (1/10th of the value range) triangular function for the opacity transfer function, altering only the peak value in the center of the window. As shown in the Figure, the model consistently provided accurate assessments in all cases. 
More details on the experiment and other assessments for other datasets can be found in the supplementary material. 

\noindent\textbf{Scatterplot.}
Compared to volume rendering images, in which visual recognition is a simple binary task, i.e., object recognition, the assessment of visual structure in scatter plots is more nuanced. 
Here, we design five basic visualization tasks: \textit{cluster recognition, cluster counting, outlier detection, outlier counting, and correlation detection} to evaluate its performance.
The evaluation result is displayed in Table~\ref{tab:scatter_plot_perceptron}.
For cluster recognition, our experiments show that the model can easily tell the plot has clusters (100\% success rate). 
However, for the counting task, the success rate of the model is only at 60\%.
Similarly, we also separate the outlier tasks into recognition and counting.
The final result aligns with the cluster recognition task.
The model performs well on the outlier recognition task which has a 100\% success rate, but has medium performance on the outlier counting task.
In the correlation detection task, the model has a 100\% success rate.


\begin{figure}[!htbp]
\centering
\includegraphics[width=0.5\textwidth]{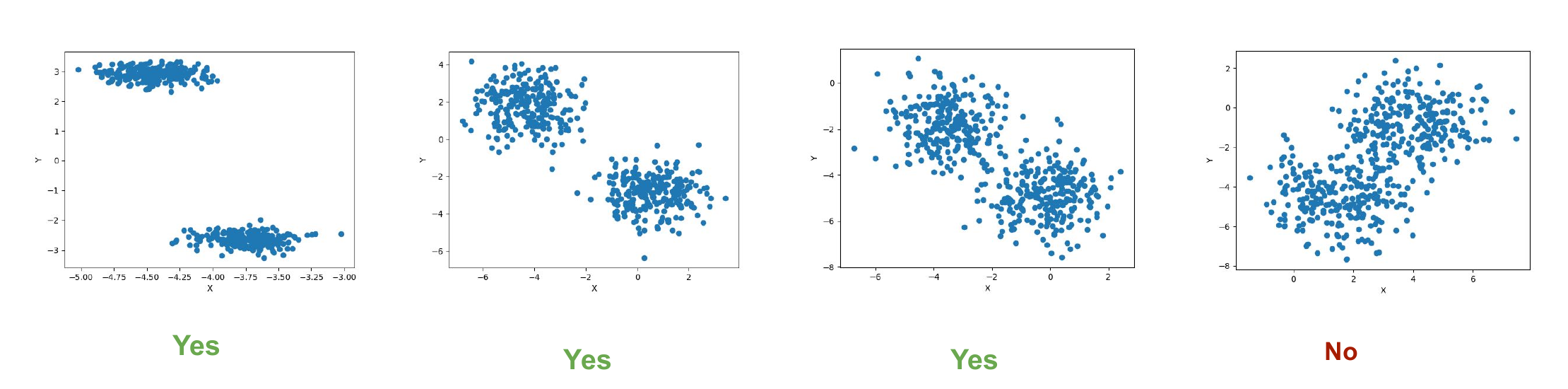}
\caption{The ability of GPT4-V to identify clusters in the scatter plot with different levels of ambiguity.}
\label{fig:scatter_plot_ambiguous}
\end{figure}

The experiment results show that the model has a decent ability to understand and analyze the scatter plot.
However, in the experiment, most of the visualizations have clear signals to tell whether certain features exist.  
This raises another question of whether the model is able to identify ambiguous cases.
We perform another simple experiment in which a scatter plot has two clusters but with different point spreads. 
From the result of Fig~\ref{fig:scatter_plot_ambiguous}, we can tell that except for the last one which it is hard for humans to tell whether it has two clusters or not, the LLM model is able to identify the rest of the example accurately. 

\begin{table}[!htbp]
\centering
\begin{tabular}{ccc}
 \hline
Tasks  & scatter plot(success rate) & parallel coordinates\\ \hline 
cluster & 100\%  & 100\%\\
cluster count & 60\%  & 20\%\\
outlier & 100\% & 90\%\\ 
outlier count & 60\% & 80\% \\
correlation &100\% & 20\% \\ \hline 
\end{tabular}
\caption{The performance of GPT4-V on a scatter plot and parallel coordinate tasks. GPT4-V can identify outlier and cluster well in both visualizations. However, its ability for object counting is comparatively poor. Meanwhile, the correlation detection in parallel coordinates plot is also limited.}
\label{tab:scatter_plot_perceptron}
\end{table}

\noindent\textbf{Parallel Coordinates Plot.}
We examine parallel coordinate plots with the same tasks as the scatter plot.
Both experiments have a similar setup on cluster and outlier tasks, except the number of dimensions in each dataset will change from 2D to 5D.
The overall results are a bit worse than the model's performance on scatterplot visualization.
In cluster and outlier recognition tasks, the model performs well.
In the cluster counting task, parallel coordinates perform badly with a 20\% success rate but in the outlier counting task, the GPT4-V model performs well.
Opposite to the correlation task, the parallel coordinate makes it hard to identify correlation relationships.

\begin{table}[!htbp]
    \centering
    \begin{tabular}{ccccc}
     \hline
     Tasks  &  node count & find node & connection & neighbor\\ \hline 
     success \% & 50\%  & 100\% & 70\% & 10\% \\ \hline 
    \end{tabular}
    \caption{The performance of GPT4-V on common graph tasks.}
    \label{tab:parallel_coordinate_perceptron}
\end{table}

\noindent\textbf{Graph.}
To assess GPT4-V's visual understanding of graphs, we choose the classic graph visualization technique node-link diagram and adjacency matrix.
In our experiment, we use the basic graph exploration task~\cite{1382886} to evaluate the performance of the LLM.
Instead of performing all tasks, we pick four tasks that are easy to perform without interactions.
The overall result is displayed in Table~\ref{tab:parallel_coordinate_perceptron}.
From the evaluation, we can tell that LLM can easily find a node in the graph visualization. However, it is difficult to tell the neighbor of the selected node. 
The connection tells whether two nodes are connected (directly or indirectly through other nodes), and the final result shows sub-optimal performance. 
Finally, the node count ability has a 50\% success rate which shows that the model again has poor performance on the counting tasks.

Despite the relatively limited exploration, our experiment demonstrates the model's capability to discern structures and objects in volume rendering results. Among the information tasks, the model achieves better performance on scatterplots compared to parallel coordinate plots or graphs. Therefore, to leverage the strength of the system, in our case study (see section \ref{sec:results}), we focus on volume rendering and scatterplot-related applications.

\begin{figure*}[t]
    \centering
    \includegraphics[width=0.9\textwidth]{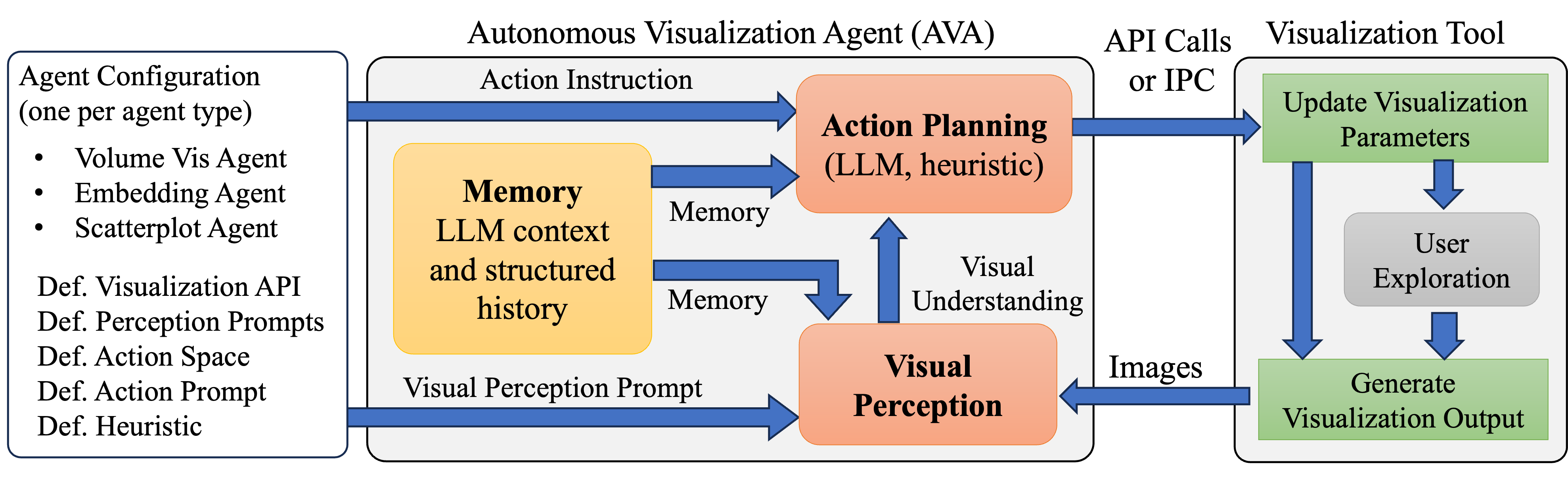}
    \vspace{-2mm}
    \caption{The breakdown of the components of the AVA. The capability of the AVAs hinged on their visual perception, and the visual understanding can then be utilized by the action planning system to modify/steer the visualization tool. In order for AVAs to make informed decisions and the ability to understand context, they also need a memory component that both visual perception and action planning components can easily access.}
    \vspace{-2mm}
    \label{fig:AVA_design}
\end{figure*}

\section{Autonomous Visualization Agent (AVA)}
\label{sec:agent_design}




We define AVA as a paradigm for designing AI-driven agents that serve as a medium between a specialized visualization tool and a domain user. The key principle of AVA involves the utilization of machine vision for decision-making. It takes user instruction in natural language and achieves the user-specified goal by operating the visualization tool autonomously based on the visual understanding of visualization outputs. And we refer to the concrete implementation of AVA as AVAs.

\subsection{Key Components of AVAs}
To achieve its design goal, the AVAs need to accurately perceive visual input and make plans on what action to take based on current visualization results and do so by following user natural language instructions. 
As illustrated in Figure \ref{fig:AVA_design}, AVAs need to contain at least three key components, namely visual perception, action planning, and memory. 

\noindent\textbf{Visual Perception} the visual perception is at the center of the AVAs' capability, and what distinguishes it from existing LLM applications in visualization. There is some similarity between AVA and an embodied agent \cite{zhang2023building} in robotic research, where an agent will take action based on sensory input (e.g., vision) and observe the impact of the action in the environment. Similarly, for AVAs the sensory input is the visualized image, and the action corresponds to changes in the visualization setting (e.g., update parameters), and the impact of the action is a new visualization output. 

\noindent\textbf{Action Planning} In order to make autonomous decisions and respond to the ``sensory'' input, the AVAs need an action planning component as the ``brain'' of the system. 
Here we have a range of choices for its design. 
As illustrated in Figure \ref{fig:agent_design_space}, we can either rely more on heuristics to drive the action planning or let the LLM do everything on its own, which corresponds to two distinct approaches to the action planning design. 
\begin{itemize}
    \item Heuristic-Centric: infuse our existing domain knowledge into heuristics for how to update the visualization tool based on assessment from the visual perception component. Their action plan is defined explicitly. In such a scenario, the visual perception and assessment essentially act as a loss function for a pre-defined optimization procedure.
    \item LLM-Centric: leverage the capability and prior knowledge of LLM to guide the exploration of the action space. Their action is only influenced by the initial prompt feed to the system.
\end{itemize}
One important thing to note is that both approaches will provide autonomous decision-making based on visual perception, so from the user's perspective there may be little difference. The distinct between them comes from whether we want to rely on our own prior knowledge explicitly or we hope to leverage the LLM's capability for planning and suggestion, while only influencing its behavior indirectly.

\noindent\textbf{Memory} Beside the visual perception, and action planning components, the other essential part of AVA that both of these components need is memory of the previous actions or the visualization outputs it observed before. In order to make complex and well-informed decisions, we often need to refer back to or compare with previously examined results or conclusions. The same is true for AVAs.

\noindent\textbf{Visualization-Perception-Action Loop} 
Besides the three key components, one essential aspect of AVA, and its key capability, is associated with the autonomous visualization loop, i.e., from visualization to perception and then to action.
The process is bootstrapped by the specific high-level task given by the user and starts with a default visualization setup, and then the system:
\begin{itemize}
    \item Generating visualization output by executing API calls to the visualization tool based on the given parameter. 
    \item Leveraging the visual perception component to comprehend semantics and structure in the current visualization.
    \item Provide assessments of whether the visualization achieves the user-set goal, and the action planning component makes decisions on what visualization parameters it should use next.
\end{itemize}

The agent will iterate through these steps until the visualization goal is achieved.
This methodological framework forms the foundation of AVAs, enabling us to utilize the visual perception system and the optimization strategy of LLMs or heuristics to interact with visualization outputs effectively. 
It addresses the critical need for autonomous agents capable of navigating complex visualization tasks with precision and adaptability.


\subsection{Implementation}
So far, we have discussed the conceptual idea of how AVA works. In this section, we provide practical guidance on their implementation.
Our implementation utilizes the GPT-4 Vision model \cite{openai2023gpt4vision} for visual perception and action planning (for LLM-centric scenario), harnessing its natural language understanding capabilities alongside a visual perception engine. 

To establish a flexible and reusable foundation for our AVAs, we have created an abstract class that acts as the basis for any specific type of agent. It contains several core functionalities: 1) a unified interface for accessing LLM API for visual perception or action planning tasks; 2) a basic blueprint on how an agent should interface with the visualization tool; 3) configuration functionality that helps define the agent, e.g., prompts template; 4) capability to parse and extract visual assessment results, parameter, and function call information from the language response.

\begin{figure}[!htbp]
    \centering
    \includegraphics[width=0.45\textwidth]{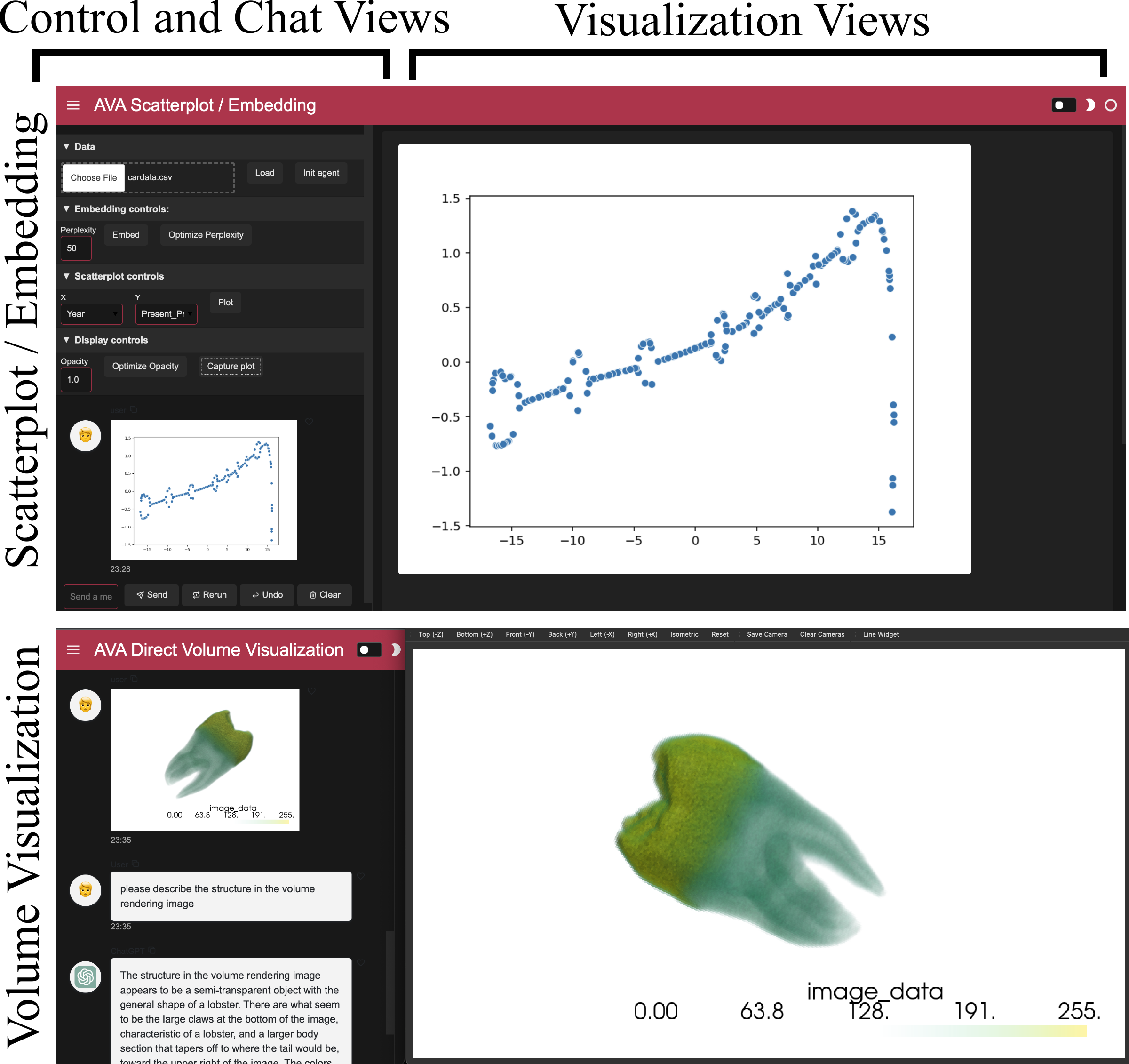}
    \caption{AVA integrated visualization interface. We implement AVA as a configurable system where individual components can be combined/configured for different target applications.}
    \label{fig:AVA_interface}
\end{figure}

For specific types of applications, AVAs can be developed as concrete classes that inherit the base one, and in the new class, application-specific logic, e.g., heuristics-centric action planning, can be implemented. Each of the concrete classes will also have an associated JSON configuration file, prompts or part of prompts are organized in a structured fashion. Our AVA implementation is in Python. As illustrated in Figure \ref{fig:AVA_interface}, we design a simple layout for the interface, with the control and conversion history on the left panel and the visualization of interests on the right.

\noindent \textbf{Agent Initialization} To initiate an AVA's functionality, we establish a context by prompting the Large Language Model (LLM) with the assumed role of the agent. This definition typically encompasses several elements: scenario, visualization task, goal, approach, and constraints. The prompt structure typically follows this format:

\noindent \textit{You are an autonomous visualization agent tasked with assisting a user in \{visualization task\}. In each step, you will receive a screenshot and you will assess the image and provide the \{approach\}. Your goal is to determine \{goal\}. Achieve this goal by \{approach\}, adhering to the following constraints: \{constraint 1, constraint 2…\}
}

One of the most critical components of AVA implementation is creating natural language prompts. Crafting effective prompts is essential for defining the agent's role and specifying the approach to achieving the visualization task. 



\noindent \textbf{Connection Between AVA and Visualization Tool}
The AVA can either directly call the API, provided both the agent and the visualization run in the same application/context. However, to maximize the flexibility and support complex external tools (e.g., GPU accelerated direct volume rendering), we also include support for a more generic solution with inter-process communication (IPC) mechanisms to facilitate seamless data exchange between the agent and the visualization tool. In our implementation, we utilize the RPyC (Remote Python Call) \cite{RPyC} to facilitate the IPC.
\begin{figure*}[t]
    \centering
    \includegraphics[width=0.99\linewidth]{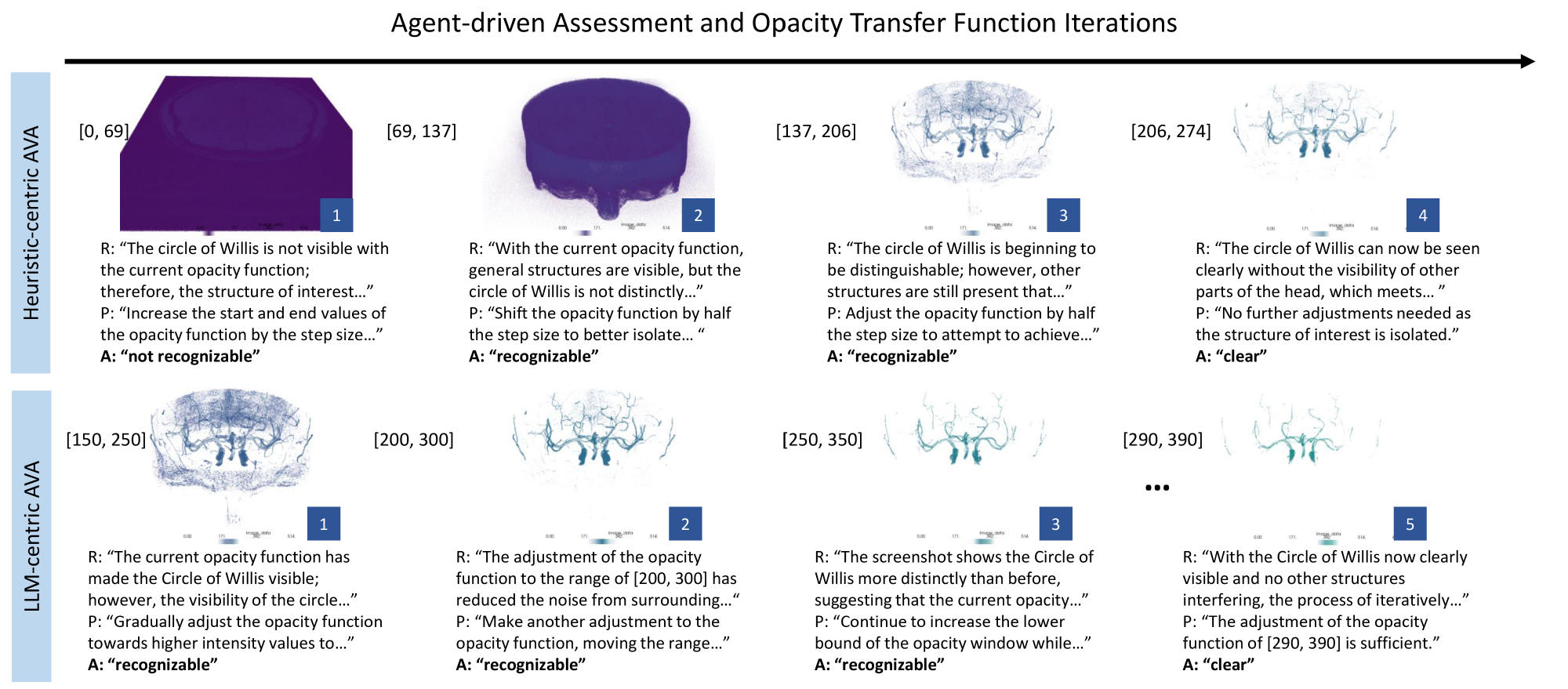}
    \caption{The results from the heuristic-centric and LLM-centric AVAs. The screenshots are generated from the proposed opacity transfer function. The response includes (R)easoning, (P)lan, and (A)ssessment by the agent, and based on this, the agent suggests a new pair of values to construct the triangle-shaped opacity transfer function. Each agent converged towards an opacity function rendering the structure of interest.}
    \label{fig:autonomy_levels}
\end{figure*}

\section{Case Studies}
\label{sec:results}
\subsection{Opacity Transfer Function for Volume Rendering}
In this case study, we focus on the opacity transfer function design process—a crucial task in volume rendering, where structures of interest must be appropriately depicted within the opaque range of the opacity transfer function. We test the agents with a dataset of a head \cite{head_dataset}, which is a 3T Time-of-flight Magnetic Resonance Angiography and contains part of the portion around the height of the eyes where the brain arteries are located. 
It contains skin, soft tissues, the skull, and the vascular structure inside. The most interesting structure inside this dataset is the arterial blood supply of the brain, called the circle of willis. 

\noindent \textit{Scenario: Agent assists the user in volume rendering.\\
Visualization Task: Evaluate the visualization output and determine the appropriate opacity transfer function for rendering a structure of interest within a volume.\\
Goal: Identify an opacity transfer function that accurately renders the structure of interest.}

To facilitate a comprehensive discussion of AVA behavior, we implemented two different agents. A heuristic-centric agent receives the action plan as a heuristic defined by the user, while the LLM-centric agent utilizes the model's knowledge about the opacity transfer function design in order to devise a strategy. 
For both agents, the opacity function remained a triangle function with the peak value positioned between the start and endpoints. The viewpoint and color map were also fixed. The AVAs provide assessments categorized as \textbf{'recognizable'}, \textbf{'not recognizable'} as described in Section 2. In addition, we added the \textbf{'clear'} assessment, as a stopping criterion for the optimization, which denotes that the structure of interest is distinctly visible without any other structures occluding it. This assessment is necessary for the AVA in order to improve results upon finding an opacity value range that renders the structure of interest 'recognizable', which could still contain a large amount of noise. The agents are described as follows (details are available in the supplementary material).

\begin{figure*}[t]
    \centering
    \includegraphics[width=0.99\linewidth]{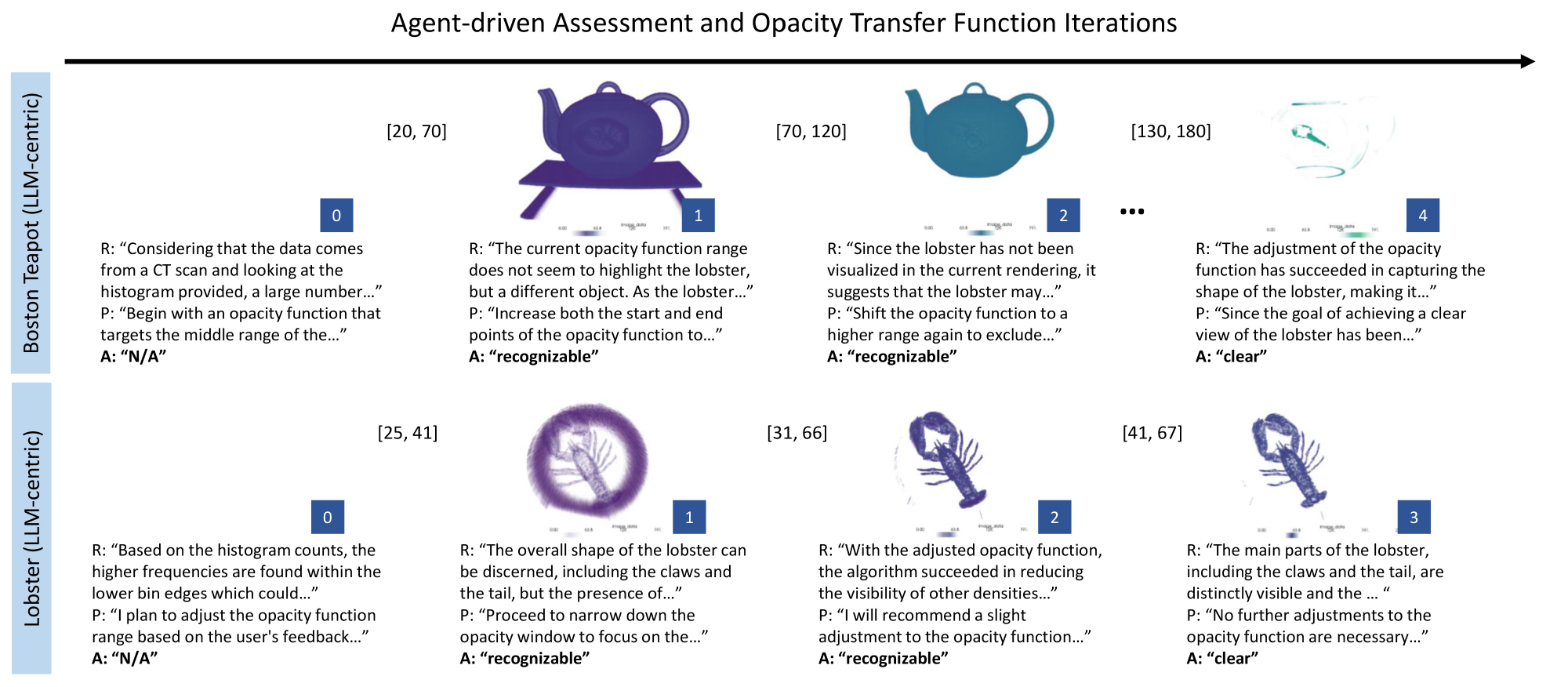}
    \caption{The results from the LLM-centric AVA for the Boston Teapot and the Lobster in Resin datasets. The structure of interest is \textit{the lobster} in both cases. The rendered images are generated from the proposed opacity transfer function. The response includes (R)easoning, (P)lan, and (A)ssessment by the agent. In the top row, the agent suggested the first opacity transfer function that revealed the teapot instead of a lobster and it moved on higher value ranges and successfully detected the lobster, even at a low resolution. In the bottom row, AVA found the lobster, which has higher resolution and occupies a larger space in the volume, and immediately and fine-tuned the result until almost no resin was visible in the visualization output.}
    \label{fig:high_autonomy}
\end{figure*}

\noindent \textbf{Heuristic-Centric:} In this setup, the agent provides assessments, but the opacity transfer function adjustments are defined by the agent designer. For a proof-of-concept, we utilize a simple linear search-based approach that shifts the window of the opaque range towards higher values, while the function always assumes a triangle shape. For these tests, we selected parameters to separate the value range into 10 bins, where the window width is one bin wide. We shift the window one bin with each iteration. We also added a fine-tune parameter, where we reduce the speed when the structure of interest is "recognizable' but it is not yet 'clear'. In that case, the window shifts only by half of its width. The only information the domain user needs to provide here is the structure of interest (the circle of Willis, a vascular structure in the brain) and the value range.  

In general, this action plan can be implemented in two ways, depending on the scenario. This heuristic can be added as code, such as a plugin integrated into the visualization code.
To elucidate the agent's behavior in alignment with the heuristic described above, the agent initially employs a triangle function at the far left end of the value range and incrementally moves upward. It continues this process until it can confidently recognize the structure of interest, in this case, the circle of Willis. As the agent recognizes the vascular structure, it takes half a step to make smaller adjustments to the opacity transfer function until it can fully discern the circle of willis. While we used a linear search strategy in this demonstration, other approaches, such as binary search, can also be employed. An advantage of utilizing large language models (LLMs) is the ability to incorporate text from research papers on opacity transfer function design, providing the agent with a more advanced action plan.

\noindent \textbf{LLM-Centric:} This AVA is not limited to a user-defined heuristic for adjusting the opacity function. Instead, it can leverage the prior knowledge of transfer function design inherent in the LLM to facilitate the design process. However, it remains constrained to providing a triangle function as the opacity transfer function.
We provided the agent with the acquisition modality and the histogram to provide it with similar information as a human user would have. 
As depicted in Figure \ref{fig:autonomy_levels}, the agent explores various opacity ranges until it successfully generates the function for the circle of willis. Notably, in this case, the agent operates with a greater degree of autonomy, employs a strategic approach, and reflects on past decisions as explained in its "reasoning" and "plan" as shown in Figure \ref{fig:autonomy_levels}. Interestingly, it immediately devised a plan, where it starts with a range higher than the first peak in the histogram, which it correctly assumes is the background. 

We further tested the AVA's capabilities on structures that are more challenging to find an appropriate opacity function. Specifically, we utilized the Boston Teapot dataset discussed in Section 4, which contains a lobster inside the teapot. The lobster in this dataset can be only partially visualized due to the low resolution of the data.
This presented a more difficult scenario compared to the circle of willis. Additionally, the lobster is relatively small within the dataset, resulting in its representation by a very low bin in the histogram.
Despite these challenges, as demonstrated in Figure \ref{fig:high_autonomy}, the agent successfully determined the correct opacity transfer function within a few steps when tasked with identifying the lobster structure of interest. Following the reasoning in each step, it reveals its advanced action-planning capabilities. To provide a comparison, we also tested the agent on another dataset, containing a Lobster in Resin (\textit{301x324x56, uint8}, Courtesy of VolVis distribution of SUNY Stony Brook, NY, USA.). In both cases, the model reasoned that the second histogram peak might be the structure of interest, however in the Boston Teapot dataset, it found the teapot instead of the lobster. Remarkably, the LLM-centric agent moved on and tried different value ranges and found the lobster in just a few steps, even though the lobster is harder to recognize due to the low resolution. In comparison, in the Lobster in Resin dataset, the lobster was revealed together with the resin in the first iteration and then the agent fine-tuned the opacity in Step 3 until no resin is visible anymore. 

This successful demonstration illustrates the agent's robustness in handling challenging scenarios and its ability to swiftly adapt to different datasets and structures of interest.
The detailed results of these agents and their responses are provided in the supplementary material.


\subsection{Scatterplot Opacity Optimization}
Apart from rendering output, from our initial assessment (section \ref{sec:vison_eval}) the GPT-V has better visual perception for scatterplot compared to all other common information visualization encodings (e.g., parallel coordinate, graph). Therefore, we focus the rest of the case studies on the scatterplot type of visual output.
In this section, we examine the optimization of opacity value for scatterplot points to mitigate the occlusion effects from overplotting. 

The perceptual base opacity optimization has been explored in the visualization domain \cite{matejka2015dynamic, micallef2017towards}, either through a data-driven modeling perspective based on user preferences \cite{matejka2015dynamic} or through a visual perception modeling approach by designing a cost function that captures relevant aspects of the human visual response \cite{micallef2017towards}. Here we do not aim to directly compare with these existing methods, as a meaningful comparison requires an extensive and controlled study. We hope to use this case study to illustrate how a fundamentally different approach to address the opacity optimization challenge can be obtained by a straightforward adoption of the AVA framework. 
From the existing study on user preference \cite{matejka2015dynamic}, the relationship between the point opacity and assessment of overplotting level follows an inverse logarithmic relationship, i.e., overplotting only gets better when point opacity gets much lower in the 0.0, 1.0 range. 
This is crucial prior knowledge that should be incorporated into the design of the agent. Therefore, we adopted the heuristic-centric approach outlined in Section \ref{sec:agent_design}, where we encode the logarithmic relationship into our search procedure. At the start of the optimization, we set the initial opacity $O = 1.0$. The floor opacity, i.e., lowest allowable opacity $O_f = 0.0$. For each step, we will update the new opacity as $O' = O_f + (O - O_f)/2$, essentially half the opacity value different between the current opacity and the floor opacity. By providing the model with scatterplot images generated with opacity $O'$ and $O$, we then evaluate which opacity is better suited for the given data. If the new opacity is deemed too low, we then set it as the new floor opacity $O_f$. We continue to iterate to narrow down the selection until the opacity different threshold is reached.

\begin{figure}[!htbp]
    \centering
    \vspace{-2mm}
    \includegraphics[width=0.5\textwidth]{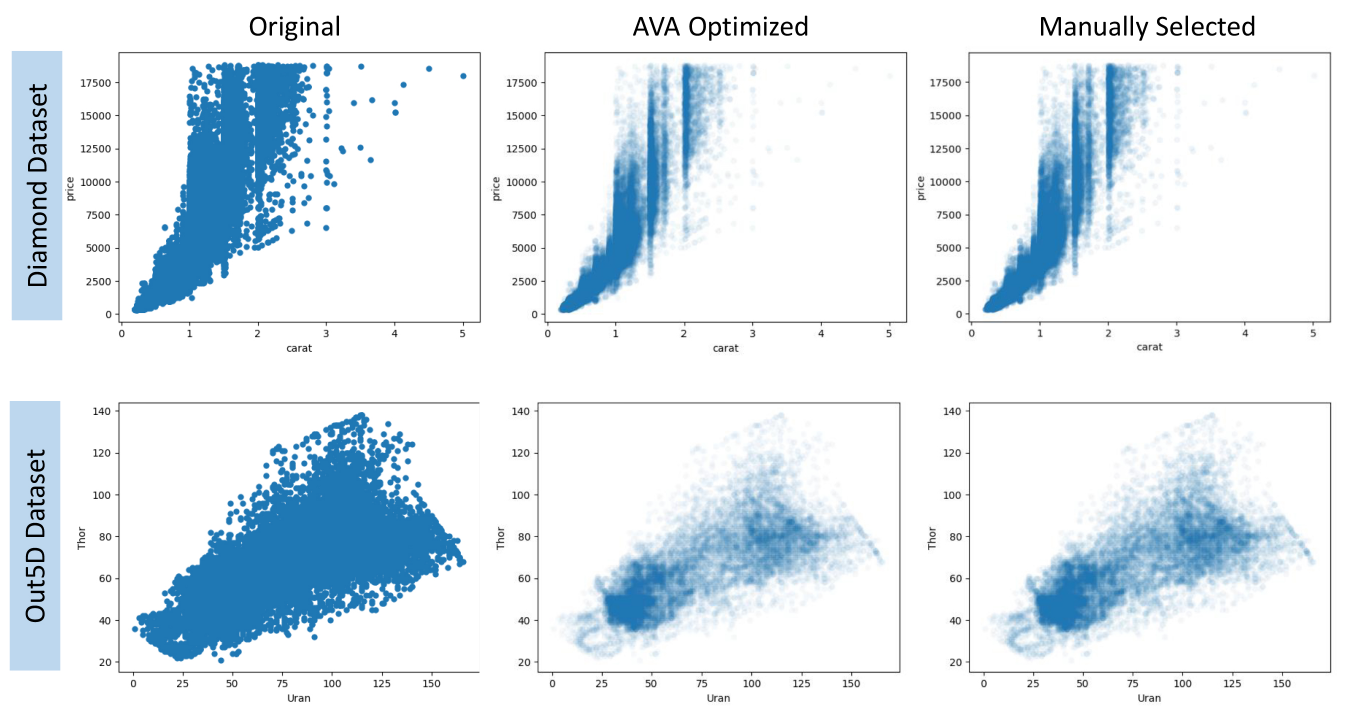}
    \vspace{-4mm}
    \caption{Scatterplot opacity optimization results. The left column shows the original plots with severe overplotting, the middle column shows the agent optimization results, and the right column shows independent manual selection results, there are some minor differences, but the overall results are comparable. }
    \label{fig:scatterplot_opacity}
    \vspace{-1mm}
\end{figure}

As shown in Figure \ref{fig:scatterplot_opacity}, each row indicates a different dataset, namely Diamond data \cite{kaggleDiamondDataset}, and the Out5D data \cite{xmdvHomepage}. The first column includes the original scatterplots with the overplotting issue when opacity is 1.0, the middle row is the AVA opacity-optimized scatterplots, and the last row is the human user reference obtained independently from the optimization interface.
As we can see, the AVA-generated scatterplot closely matched the user preference. 




\subsection{Dimension Reduction Hyperparameter Tuning}
\begin{figure}[!htbp]
    \centering
    \vspace{-2mm}
    \includegraphics[width=0.45\textwidth]{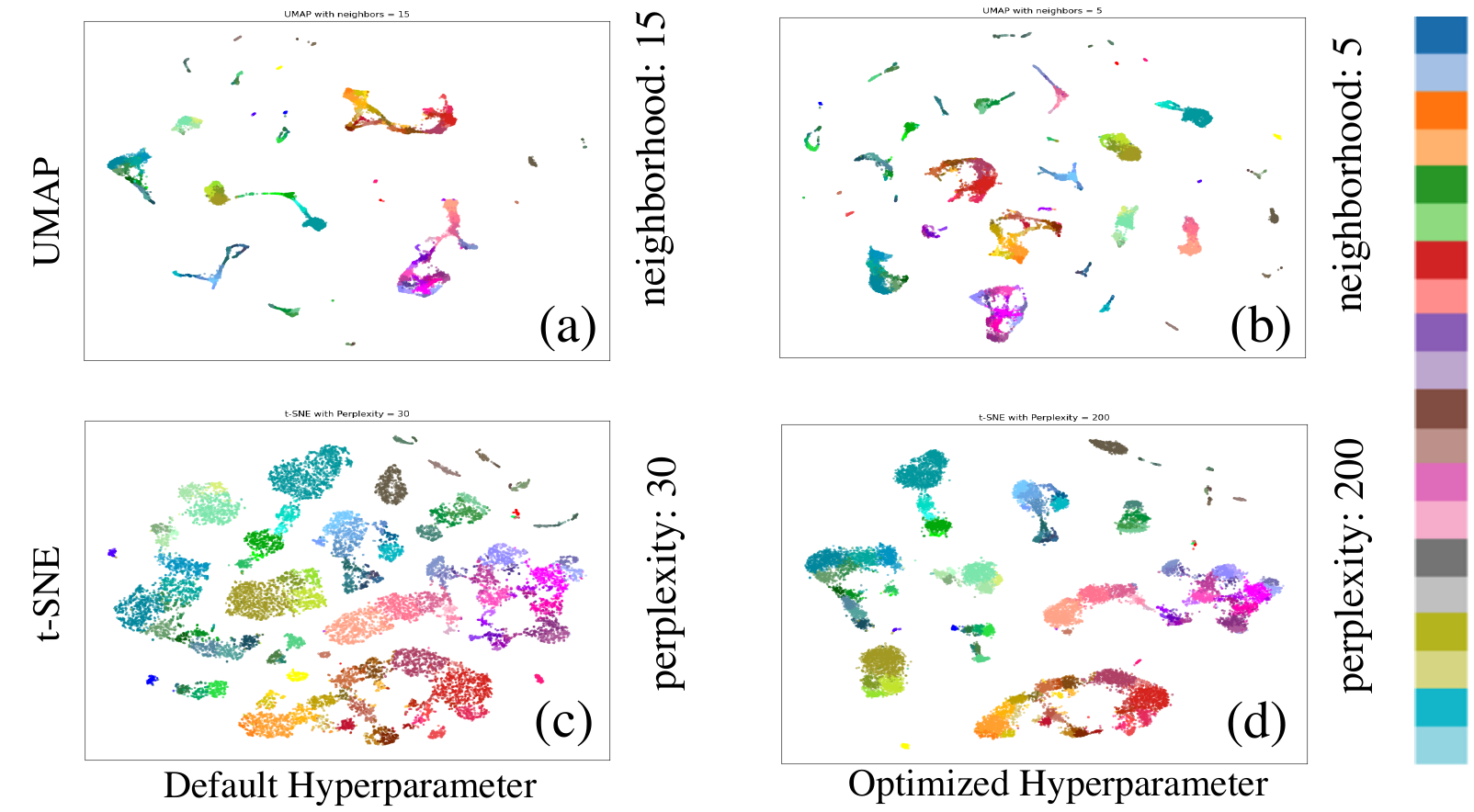}
    \caption{Hyperparameter optimization results. For UMAP, the optimized hyperparameter (b) shows greater class separation compared to the default (a). For t-SNE, (d) shows a similar class separation but with more compact clusters.}
    \label{fig:embedding}
    \vspace{-2mm}
\end{figure}

The choice of hyperparameters can greatly impact t-SNE \cite{van2008visualizing} and UMAP \cite{mcinnes2018umap} results. Inappropriate hyperparameters may lead to misleading interpretations of the high-dimensional structure, and they often need to be tuned for a given dataset. Here, we utilize AVA to perform automatic hyperparameter tunning for identifying more suitable hyperparameters, for both single-hyperparameter and multi-hyperparameter cases. Considering the prior knowledge the LLM is likely to have on these common methods, we opt for LLM-centric action planning, where the LLM directly suggests hyperparameters.
In Figure \ref{fig:embedding}, we show the single parameter optimization result, where we only optimize the most sensitive parameter for each method, i.e., perplexity for t-SNE, and the neighborhood size for UMAP. We withheld the class label from the agent to use as the ground truth for evaluation.
All plots are generated from the data RNA sequence data \cite{tasic2018shared} with 20 classes. As we can see for the UMAP embedding, the default parameter gives a small number of stringy clusters (a), whereas, in the optimized embedding (b), several classes that were linked together are now separated.
For the t-SNE case, there is a less clear advantage for the optimized embedding in terms of cluster separability, however, it does show more compact clusters.

Our experiment with the multi-hyperparameter(up to 5) agent case, however, is largely unsuccessful. As we see the suggested hyperparameters bounce back and forth during the optimization process. This indicates the potential challenge for the agent to explore higher dimensional action space (see more detail in Section \ref{sec:discussion}).

\section{Experts Feedback on AVA}
Given that we did not aim to conduct a general evaluation of multi-modal foundation models, nor did we target a specific application area, we chose to gather informal feedback from experts in key domains that could provide valuable insights into the potential of these agents for visualization tasks. Our feedback collection process involved two senior AI researchers, a professor in medical visualization, and a professor who heads an Institute for Radiological Diagnostics and Intervention. 
The latter two were selected due to their daily workflow experience with volumetric data and general visualization tasks. 
For all the feedback sessions, we first demonstrated the use cases of AVAs as described in Section \ref{sec:results}, and then conducted unstructured interviews with the experts. While the discussions were mostly open-ended, we did inquire about their overall feedback on how such visualization agents might impact their workflow, and where the potential benefits and limitations of this paradigm are. 

\noindent \textbf{The medical visualization expert said:}
\textit{
``Their ability to comprehend visual elements and identify structures is indeed impressive, laying a solid foundation for the future development of such agents. With this substantial potential at hand, the utilization of these agents now lies in the hands of visualization researchers, who have the opportunity to harness their capabilities for innovative applications.''}
She extended the discussion by suggesting the creation of a generic workflow that can incorporate how visualization experts use the volume rendering tool into the agent prompts to enhance their capabilities. 

\noindent \textbf{The head of the radiology institute said:}
\textit{``I'm impressed with the semantic understanding, reasoning capabilities, and high autonomy exhibited by the agents. There is exciting potential to replace trivial visualization tasks that until today require a radiologist.''}
He envisioned the use of these agents for double reading in radiology, where two independent radiology reports could be generated to cross-validate diagnoses. 
However, he remained skeptical about whether the recognition could go beyond simple shapes, for example actually perceiving differences between arteries and veins, as well as extending the visual perception capability to make assessments based on multiple image modalities.

\noindent \textbf{The senior AI researchers said:}
\textit{``1: This can be a very general approach. One additional application I can see this working is for finding more informative views for 3D plots, which I always have trouble with.''}  \textit{``2: The AVA setup can be easily extended to other types of user interfaces beyond just visualization. One thing I am interested to know is how well it handles a larger action space, will the search fail or converge?''}
The AI experts believe this is a fundamentally different way to think about data visualization problems and see the connection with embodied agent research. One potential concern they mentioned is whether the action planning can work with a much bigger action space. As a response to their feedback, we extend our dimensionality reduction case studies to include additional experiments with up to 5D space.
Overall, the feedback from all the experts underscores the transformative potential of AVA. 
\section{Discussion and Future Work}
\label{sec:discussion}


In this work, we investigated the capabilities of emerging multi-modal foundation models like GPT-4V and their ability for visualization tasks. 
Building on these findings, we introduced a template for a novel paradigm known as Autonomous Visualization Agent (AVA) for solving high-level visualization tasks through visual perception and action planning.
Despite demonstrating its feasibility through our case studies, it
is also essential to acknowledge their limitations. 

\noindent\textbf{Prompts Engineering.} With the flexibility and usability of natural language, it also brings certain limitations. The precision of AVAs heavily relies on the choice of the prompt, as the natural language remains inherently fuzzy and context-dependent. This limitation may diminish in significance as these models evolve to become more powerful and context-aware. 
This challenge can also be partially mitigated through heuristic-centric action planning through explicitly coded search logic, nevertheless, the visual perception still relies on prompts to convey the assessment objectives.

\noindent\textbf{Large Action Space.} The action planning component of AVA is essentially doing an exploration of a potentially high-dimensional action space. The search is guided by the visual assessment, which theoretically can be considered as the loss in a zeroth order optimization scenario \cite{slavin2022adapting}. However, despite having a black box loss, we as visualization designers or the LLM do have prior knowledge of the action space that could guide the exploration to allow a fast convergence. Still, as the problem is associated with the degree of freedom of the underlying system, there is no easy solution, we believe LLM model with stronger prior (i.e., visualization task fine-tuned models) and better optimization strategy is likely required to allow AVAs to reach their full potential.

\noindent\textbf{Future Directions.} Our plan for future work involves a more extensive evaluation of the current models' capabilities in understanding visualization output, expanding on the foundation laid in Section 4. This evaluation will provide deeper insights into the extent to which multi-modal foundation models can contribute to visualization research.
Additionally, we plan to explore different agent setups, including increasing the number of agents and increasing the interactivity of the agent. By implementing multiple independent agents with slightly different definitions, we can offer a means of cross-validation for applications with low error tolerances.
So far, we have demonstrated agents employing a closed-loop optimization strategy with intermittent communication with the user. 
By tuning the level of interactivity, as in a chatbot, we could create an even tighter symbiosis between a human expert and an AI for a joint visualization task.  
For additional application scenarios. There are many possibilities, as mentioned by one expert we interviewed, adjusting the viewpoint to avoid visual occlusion in 3D visualization or 3D plots can be a great use case. They can be particularly useful for offline rendering in HPC applications and large-scale data etc. The model's ability for scatterplot understanding can also be utilized to design customized diagnostics metrics for exploratory data analysis.

\section{Conclusion}
The primary objective of this work is to underscore the significance of autonomous visualization agents in enhancing the accessibility of visualization tools. We have demonstrated that not only are these agents possible, but they can already be useful for solving non-trivial visualization tasks. As multi-modal foundation models continue to advance in power and sophistication, we anticipate a corresponding increase in the capabilities of such agents.
In many ways, we are speculating on how the ongoing development of Large Language Models (LLMs) can reshape the landscape of visualization research. The fusion of image understanding and language understanding within these multi-modal foundation models holds the potential to fundamentally transform the way we think about visualization and user interaction. 
With further development, we believe AVAs can eventually serve as virtual visualization experts in the room, streamlining the entire visualization pipeline beyond the automatic visualization parameter adjustments that have been demonstrated in this paper.
In conclusion, our research opens exciting possibilities for the future of visualization tool design that aims at the collaboration between humans and AI-driven agents. 
\section*{Acknowledgement}
This work was performed under the auspices of the U.S. Department of Energy by Lawrence Livermore National Laboratory under Contract DE-AC52-07NA27344. The project is supported by LLNL LDRD (23-ERD-029, 23-SI-003). The work is reviewed and released under LLNL-CONF-857838. We thank the experts Prof. Renata Raidou (TU Wien), Prof. Christian Nasel (Medical University Vienna), Dr. Jayaraman Thiagarajan (LLNL), and Dr. 
Rushil Anirudh (LLNL) for their valuable feedback on the capabilities of our agents. 
\bibliographystyle{eg-alpha-doi} 
\bibliography{refs}       

\newcommand{\etalchar}[1]{$^{#1}$}
\begin{thebibliography}{\uppercase{MPOW17}}

\bibitem[CZW{\etalchar{*}}23]{chen2023beyond}
\textsc{Chen Z., Zhang C., Wang Q., Troidl J., Warchol S., Beyer J., Gehlenborg
  N., Pfister H.}:
\newblock Beyond generating code: Evaluating gpt on a data visualization
  course.
\newblock \emph{arXiv preprint arXiv:2306.02914} (2023).

\bibitem[DD19]{dibia2019data2vis}
\textsc{Dibia V., Demiralp {\c{C}}.}:
\newblock Data2vis: Automatic generation of data visualizations using
  sequence-to-sequence recurrent neural networks.
\newblock \emph{IEEE computer graphics and applications 39}, 5 (2019), 33--46.

\bibitem[Dib23]{dibia2023lida}
\textsc{Dibia V.}:
\newblock {LIDA}: A tool for automatic generation of grammar-agnostic
  visualizations and infographics using large language models.
\newblock In \emph{Proceedings of the 61st Annual Meeting of the Association
  for Computational Linguistics (Volume 3: System Demonstrations)} (Toronto,
  Canada, July 2023), Association for Computational Linguistics, pp.~113--126.
\newblock URL: \url{https://aclanthology.org/2023.acl-demo.11}, \href
  {https://doi.org/10.18653/v1/2023.acl-demo.11}
  {\path{doi:10.18653/v1/2023.acl-demo.11}}.

\bibitem[FAT99]{fujishiro1999automating}
\textsc{Fujishiro I., Azuma T., Takeshima Y.}:
\newblock Automating transfer function design for comprehensible volume
  rendering based on 3d field topology analysis.
\newblock In \emph{Proceedings Visualization'99 (Cat. No. 99CB37067)} (1999),
  IEEE, pp.~467--563.

\bibitem[Fil13]{RPyC}
\textsc{Filiba T.}:
\newblock Rpyc (remote python call), 2013.
\newblock Python library for remote procedure calls.
\newblock URL: \url{https://rpyc.readthedocs.io/en/latest/}.

\bibitem[FZF{\etalchar{*}}23]{feng2023layoutgpt}
\textsc{Feng W., Zhu W., Fu T.-j., Jampani V., Akula A., He X., Basu S., Wang
  X.~E., Wang W.~Y.}:
\newblock Layoutgpt: Compositional visual planning and generation with large
  language models.
\newblock \emph{arXiv preprint arXiv:2305.15393} (2023).

\bibitem[GFC04]{1382886}
\textsc{Ghoniem M., Fekete J.-D., Castagliola P.}:
\newblock A comparison of the readability of graphs using node-link and
  matrix-based representations.
\newblock In \emph{IEEE Symposium on Information Visualization} (2004),
  pp.~17--24.
\newblock \href {https://doi.org/10.1109/INFVIS.2004.1}
  {\path{doi:10.1109/INFVIS.2004.1}}.

\bibitem[hea]{head_dataset}
Head dataset.
\newblock \url{http://www.celebisoftware.com/Dataset.aspx?catId=3}.
\newblock Accessed: [Your Access Date Here].

\bibitem[Hun07]{hunter2007matplotlib}
\textsc{Hunter J.~D.}:
\newblock Matplotlib: A 2d graphics environment.
\newblock \emph{Computing in science \& engineering 9}, 03 (2007), 90--95.

\bibitem[kag]{kaggleDiamondDataset}
Kaggle diamond dataset.
\newblock \url{https://www.kaggle.com/datasets/shivam2503/diamonds/data}.
\newblock Accessed: YYYY-MM-DD.

\bibitem[LKG{\etalchar{*}}16]{ljung2016state}
\textsc{Ljung P., Kr{\"u}ger J., Groller E., Hadwiger M., Hansen C.~D.,
  Ynnerman A.}:
\newblock State of the art in transfer functions for direct volume rendering.
\newblock In \emph{Computer graphics forum} (2016), vol.~35, Wiley Online
  Library, pp.~669--691.

\bibitem[LWZ{\etalchar{*}}21]{li2021kg4vis}
\textsc{Li H., Wang Y., Zhang S., Song Y., Qu H.}:
\newblock Kg4vis: A knowledge graph-based approach for visualization
  recommendation.
\newblock \emph{IEEE Transactions on Visualization and Computer Graphics 28}, 1
  (2021), 195--205.

\bibitem[MAF15]{matejka2015dynamic}
\textsc{Matejka J., Anderson F., Fitzmaurice G.}:
\newblock Dynamic opacity optimization for scatter plots.
\newblock In \emph{Proceedings of the 33rd Annual ACM Conference on Human
  Factors in Computing Systems} (2015), pp.~2707--2710.

\bibitem[MHSG18]{mcinnes2018umap}
\textsc{McInnes L., Healy J., Saul N., Gro{\ss}berger L.}:
\newblock Umap: Uniform manifold approximation and projection.
\newblock \emph{Journal of Open Source Software 3}, 29 (2018), 861.

\bibitem[MNES22]{mitra2022facilitating}
\textsc{Mitra R., Narechania A., Endert A., Stasko J.}:
\newblock Facilitating conversational interaction in natural language
  interfaces for visualization.
\newblock In \emph{2022 IEEE Visualization and Visual Analytics (VIS)} (2022),
  IEEE, pp.~6--10.

\bibitem[MPOW17]{micallef2017towards}
\textsc{Micallef L., Palmas G., Oulasvirta A., Weinkauf T.}:
\newblock Towards perceptual optimization of the visual design of scatterplots.
\newblock \emph{IEEE transactions on visualization and computer graphics 23}, 6
  (2017), 1588--1599.

\bibitem[NSS20]{narechania2020nl4dv}
\textsc{{Narechania} A., {Srinivasan} A., {Stasko} J.}:
\newblock {NL4DV}: A {Toolkit} for generating {Analytic Specifications} for
  {Data Visualization} from {Natural Language} queries.
\newblock \emph{IEEE Transactions on Visualization and Computer Graphics
  (TVCG)} (2020).
\newblock \href {https://doi.org/10.1109/TVCG.2020.3030378}
  {\path{doi:10.1109/TVCG.2020.3030378}}.

\bibitem[{Ope}23]{openai2023gpt4vision}
\textsc{{OpenAI}}:
\newblock Gpt-4 vision.
\newblock \url{https://openai.com/research/gpt-4v-system-card}, 2023.
\newblock Accessed: [Insert date of access here].

\bibitem[QLY{\etalchar{*}}23]{qin2023toolllm}
\textsc{Qin Y., Liang S., Ye Y., Zhu K., Yan L., Lu Y., Lin Y., Cong X., Tang
  X., Qian B., et~al.}:
\newblock Toolllm: Facilitating large language models to master 16000+
  real-world apis.
\newblock \emph{arXiv preprint arXiv:2307.16789} (2023).

\bibitem[RBL{\etalchar{*}}22]{rombach2022high}
\textsc{Rombach R., Blattmann A., Lorenz D., Esser P., Ommer B.}:
\newblock High-resolution image synthesis with latent diffusion models.
\newblock In \emph{Proceedings of the IEEE/CVF conference on computer vision
  and pattern recognition} (2022), pp.~10684--10695.

\bibitem[RDN{\etalchar{*}}22]{ramesh2022hierarchical}
\textsc{Ramesh A., Dhariwal P., Nichol A., Chu C., Chen M.}:
\newblock Hierarchical text-conditional image generation with clip latents.
\newblock \emph{arXiv preprint arXiv:2204.06125 1}, 2 (2022), 3.

\bibitem[RKH{\etalchar{*}}21]{radford2021learning}
\textsc{Radford A., Kim J.~W., Hallacy C., Ramesh A., Goh G., Agarwal S.,
  Sastry G., Askell A., Mishkin P., Clark J., et~al.}:
\newblock Learning transferable visual models from natural language
  supervision.
\newblock In \emph{International conference on machine learning} (2021), PMLR,
  pp.~8748--8763.

\bibitem[SASW96]{spitzer1996visible}
\textsc{Spitzer V., Ackerman M.~J., Scherzinger A.~L., Whitlock D.}:
\newblock The visible human male: a technical report.
\newblock \emph{Journal of the American Medical Informatics Association 3}, 2
  (1996), 118--130.

\bibitem[SK19]{sullivan2019pyvista}
\textsc{Sullivan C., Kaszynski A.}:
\newblock Pyvista: 3d plotting and mesh analysis through a streamlined
  interface for the visualization toolkit (vtk).
\newblock \emph{Journal of Open Source Software 4}, 37 (2019), 1450.

\bibitem[SM22]{slavin2022adapting}
\textsc{Slavin I., McKenzie D.}:
\newblock Adapting zeroth order algorithms for comparison-based optimization.
\newblock \emph{arXiv preprint arXiv:2210.05824} (2022).

\bibitem[SMV23]{suris2023vipergpt}
\textsc{Sur{\'\i}s D., Menon S., Vondrick C.}:
\newblock Vipergpt: Visual inference via python execution for reasoning.
\newblock \emph{arXiv preprint arXiv:2303.08128} (2023).

\bibitem[SMWH17]{2017-vega-lite}
\textsc{Satyanarayan A., Moritz D., Wongsuphasawat K., Heer J.}:
\newblock Vega-lite: A grammar of interactive graphics.
\newblock \emph{{IEEE} Transactions on Visualization \& Computer Graphics
  (Proc. InfoVis)} (2017).
\newblock URL: \url{http://idl.cs.washington.edu/papers/vega-lite}, \href
  {https://doi.org/10.1109/tvcg.2016.2599030}
  {\path{doi:10.1109/tvcg.2016.2599030}}.

\bibitem[SWW{\etalchar{*}}23]{song2023llm}
\textsc{Song C.~H., Wu J., Washington C., Sadler B.~M., Chao W.-L., Su Y.}:
\newblock Llm-planner: Few-shot grounded planning for embodied agents with
  large language models.
\newblock In \emph{Proceedings of the IEEE/CVF International Conference on
  Computer Vision} (2023), pp.~2998--3009.

\bibitem[TIH]{teapot}
\textsc{Terarecon~Inc MERL B., Hospital W.}:
\newblock Teapot dataset.
\newblock
  \url{http://www.gris.uni-tuebingen.de/areas/scivis/volren/datasets/data/BostonTeapot.raw.gz}.
\newblock Accessed: [Your Access Date Here].

\bibitem[TYG{\etalchar{*}}18]{tasic2018shared}
\textsc{Tasic B., Yao Z., Graybuck L.~T., Smith K.~A., Nguyen T.~N.,
  Bertagnolli D., Goldy J., Garren E., Economo M.~N., Viswanathan S., et~al.}:
\newblock Shared and distinct transcriptomic cell types across neocortical
  areas.
\newblock \emph{Nature 563}, 7729 (2018), 72--78.

\bibitem[VdMH08]{van2008visualizing}
\textsc{Van~der Maaten L., Hinton G.}:
\newblock Visualizing data using t-sne.
\newblock \emph{Journal of machine learning research 9}, 11 (2008).

\bibitem[WBZ{\etalchar{*}}23]{wu2023autogen}
\textsc{Wu Q., Bansal G., Zhang J., Wu Y., Zhang S., Zhu E., Li B., Jiang L.,
  Zhang X., Wang C.}:
\newblock Autogen: Enabling next-gen llm applications via multi-agent
  conversation framework.
\newblock \emph{arXiv preprint arXiv:2308.08155} (2023).

\bibitem[WMF{\etalchar{*}}23]{wang2023survey}
\textsc{Wang L., Ma C., Feng X., Zhang Z., Yang H., Zhang J., Chen Z., Tang J.,
  Chen X., Lin Y., et~al.}:
\newblock A survey on large language model based autonomous agents.
\newblock \emph{arXiv preprint arXiv:2308.11432} (2023).

\bibitem[WWS{\etalchar{*}}22]{wei2022chain}
\textsc{Wei J., Wang X., Schuurmans D., Bosma M., Xia F., Chi E., Le Q.~V.,
  Zhou D., et~al.}:
\newblock Chain-of-thought prompting elicits reasoning in large language
  models.
\newblock \emph{Advances in Neural Information Processing Systems 35} (2022),
  24824--24837.

\bibitem[WXJ{\etalchar{*}}23]{wang2023voyager}
\textsc{Wang G., Xie Y., Jiang Y., Mandlekar A., Xiao C., Zhu Y., Fan L.,
  Anandkumar A.}:
\newblock Voyager: An open-ended embodied agent with large language models.
\newblock \emph{arXiv preprint arXiv: Arxiv-2305.16291} (2023).

\bibitem[WZW{\etalchar{*}}23]{wang2023llm4vis}
\textsc{Wang L., Zhang S., Wang Y., Lim E.-P., Wang Y.}:
\newblock Llm4vis: Explainable visualization recommendation using chatgpt.
\newblock \emph{arXiv preprint arXiv:2310.07652} (2023).

\bibitem[xmd]{xmdvHomepage}
Xmdvtool homepage.
\newblock \url{http://davis.wpi.edu/xmdv/datasets.html}.
\newblock Accessed: [Access Date].

\bibitem[YFZ{\etalchar{*}}23]{yin2023survey}
\textsc{Yin S., Fu C., Zhao S., Li K., Sun X., Xu T., Chen E.}:
\newblock A survey on multimodal large language models.
\newblock \emph{arXiv preprint arXiv:2306.13549} (2023).

\bibitem[YLL{\etalchar{*}}23]{yang2023dawn}
\textsc{Yang Z., Li L., Lin K., Wang J., Lin C.-C., Liu Z., Wang L.}:
\newblock The dawn of lmms: Preliminary explorations with gpt-4v (ision).
\newblock \emph{arXiv preprint arXiv:2309.17421 9} (2023).

\bibitem[YWV{\etalchar{*}}22]{yu2022coca}
\textsc{Yu J., Wang Z., Vasudevan V., Yeung L., Seyedhosseini M., Wu Y.}:
\newblock Coca: Contrastive captioners are image-text foundation models.
\newblock \emph{arXiv preprint arXiv:2205.01917} (2022).

\bibitem[YZY{\etalchar{*}}22]{yao2022react}
\textsc{Yao S., Zhao J., Yu D., Du N., Shafran I., Narasimhan K., Cao Y.}:
\newblock React: Synergizing reasoning and acting in language models.
\newblock \emph{arXiv preprint arXiv:2210.03629} (2022).

\bibitem[ZDS{\etalchar{*}}23]{zhang2023building}
\textsc{Zhang H., Du W., Shan J., Zhou Q., Du Y., Tenenbaum J.~B., Shu T., Gan
  C.}:
\newblock Building cooperative embodied agents modularly with large language
  models.
\newblock \emph{arXiv preprint arXiv:2307.02485} (2023).

\bibitem[ZLL{\etalchar{*}}22]{zong2022rich}
\textsc{Zong J., Lee C., Lundgard A., Jang J., Hajas D., Satyanarayan A.}:
\newblock Rich screen reader experiences for accessible data visualization.
\newblock In \emph{Computer Graphics Forum} (2022), vol.~41, Wiley Online
  Library, pp.~15--27.

\bibitem[ZSC{\etalchar{*}}23]{zhang2023gpt4roi}
\textsc{Zhang S., Sun P., Chen S., Xiao M., Shao W., Zhang W., Chen K., Luo
  P.}:
\newblock Gpt4roi: Instruction tuning large language model on
  region-of-interest.
\newblock \emph{arXiv preprint arXiv:2307.03601} (2023).

\bibitem[ZWLQ23]{zhang2023adavis}
\textsc{Zhang S., Wang Y., Li H., Qu H.}:
\newblock : Adaptive and explainable visualization recommendation for tabular
  data.
\newblock \emph{IEEE Transactions on Visualization and Computer Graphics}
  (2023).

\end{thebibliography}


\appendix

\section{Static Visualization Assessment Experiment Setup}

\subsection{Scatter plot}
In scatter plot experiments, each task will be performed in 10 experiments, and the final results are aggregated as a percentage number (success rate).
In each experiment, 500 points are generated randomly following a pre-defined pattern (e.g., number of clusters).

\begin{itemize}
    \item \textbf{Clustering Count:}
    For the clustering task, we randomly generate 2 to 10 clusters for each task.
    
    \textit{\textbf{Prompts}: "You are a scatter plot visualization expert. Is there any cluster in this visualization? Can you tell me how many clusters are in this visualization?"}

    \item \textbf{Outlier Count:}
    Different from the cluster recognition task, the scatter plot for the outlier detection will sample 1-5 outlier points without overlap in the visualization.
    
    \textit{\textbf{Prompts}: "You are a scatter plot visualization expert. Is there any outlier in this visualization? Can you tell me how many outliers are in this visualization?"}

    \item \textbf{Correlation Detection} 
    For the correlation detection task, we randomly generated two scatter plot visualizations with different correlation efficient scores ranging from 0.1 to 1.0. 
    It is worth noticing in the experiences that if the correlation in both images is very low (e.g., 0.1, 0.2) and LLM can not distinguish which one has a higher correlation but indicates both scatter plots have a low correlation, we will count this prediction correct.
    
    \textit{\textbf{Prompts}: "You are a scatter plot visualization expert. which images have a high correlation?}
\end{itemize}

\begin{figure*}[t]
\centering
    \includegraphics[width=0.19\linewidth]{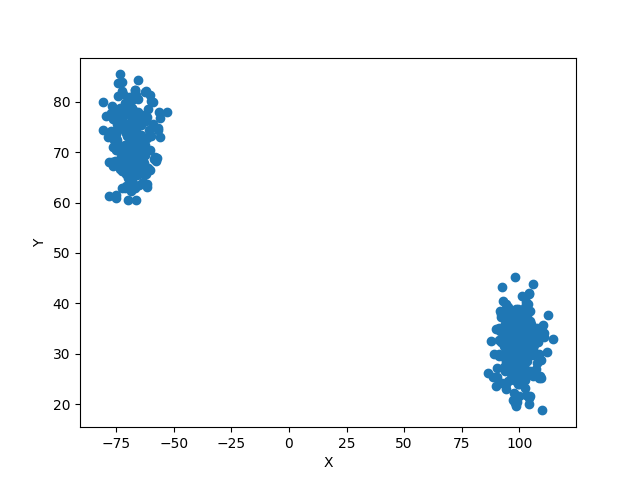}
    \includegraphics[width=0.19\linewidth]{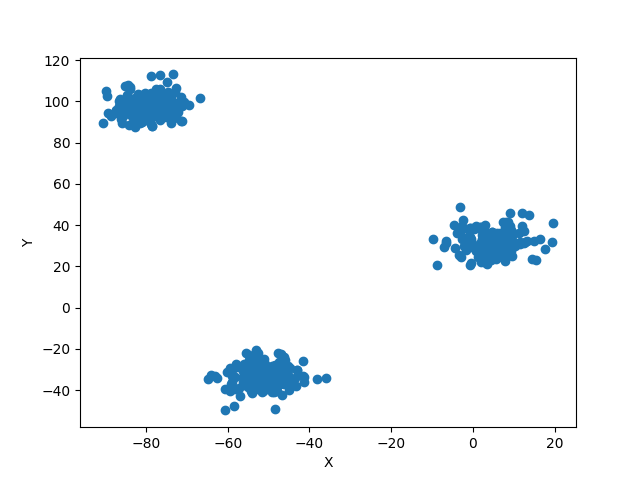}
    \includegraphics[width=0.19\linewidth]{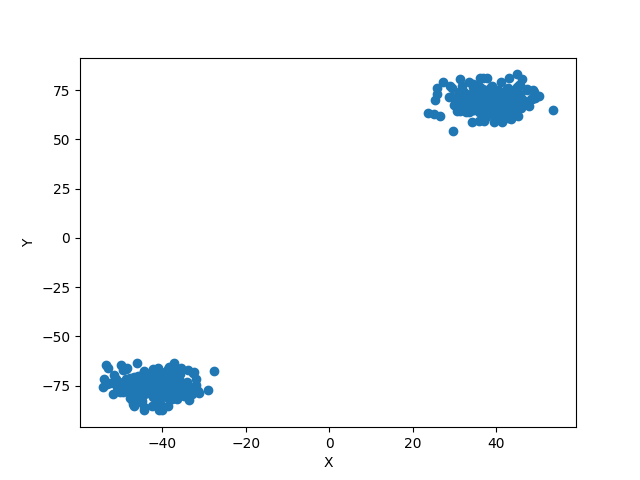}
    \includegraphics[width=0.19\linewidth]{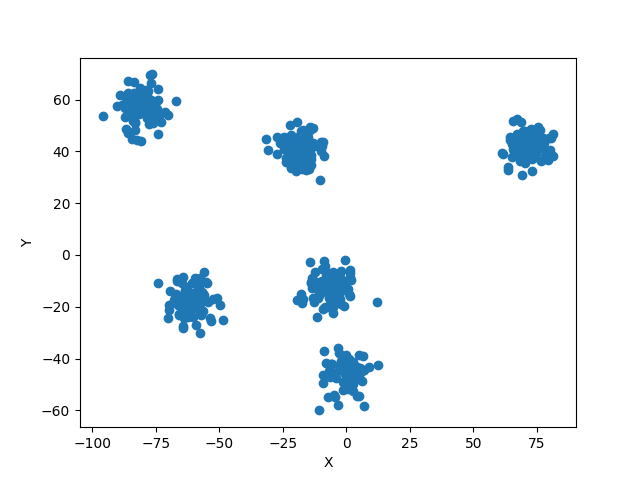}
    \includegraphics[width=0.19\linewidth]{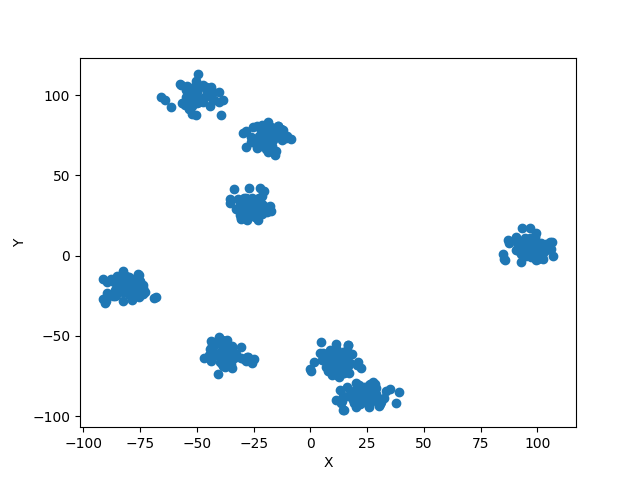}
    \includegraphics[width=0.19\linewidth]{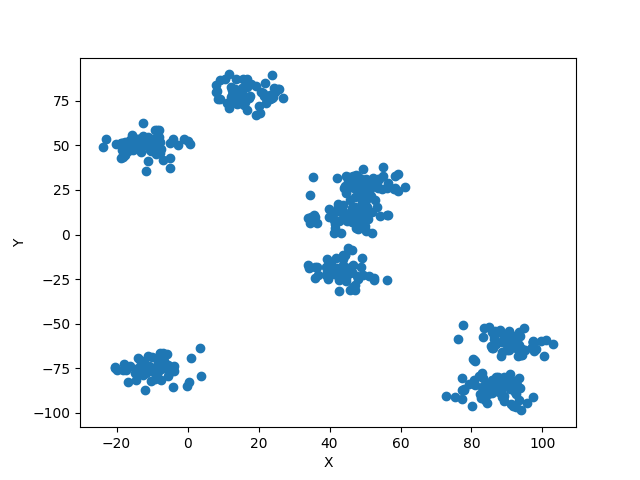}
    \includegraphics[width=0.19\linewidth]{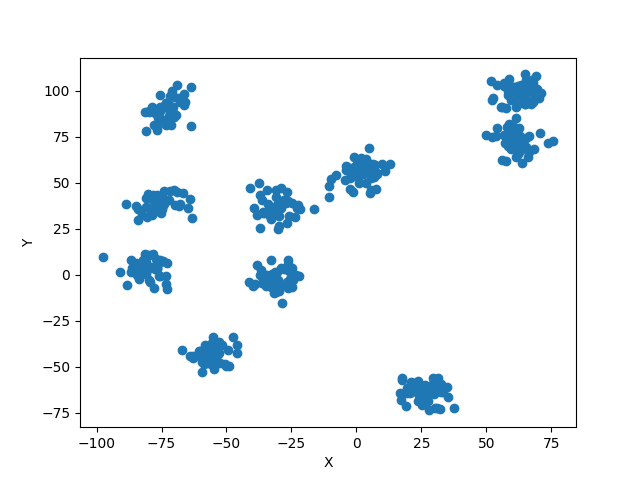}
    \includegraphics[width=0.19\linewidth]{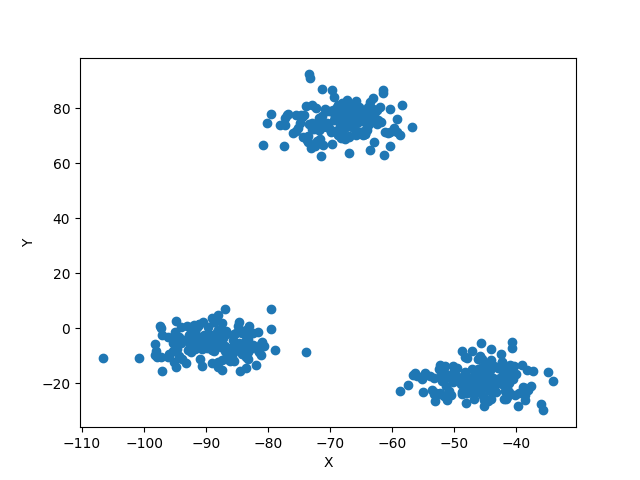}
    \includegraphics[width=0.19\linewidth]{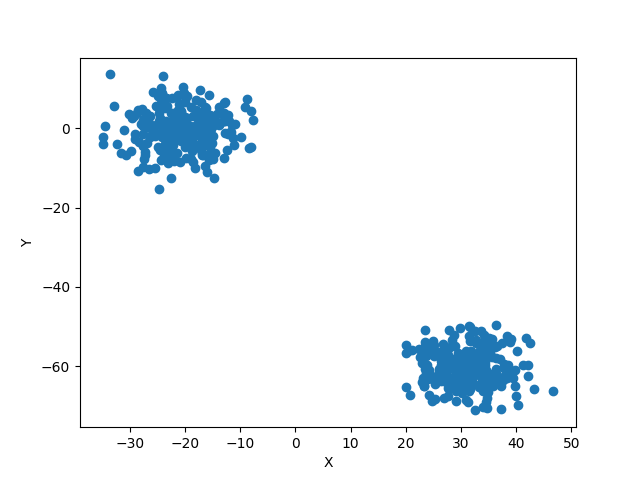}
    \includegraphics[width=0.19\linewidth]{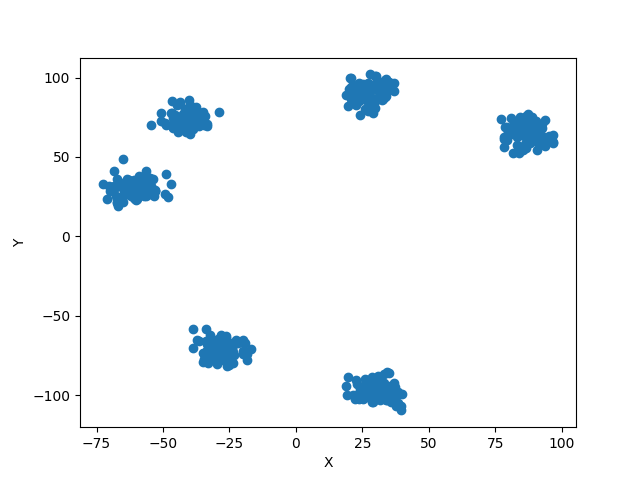}

    \caption{The randomly generated scatter plot with different numbers of clusters for LLM evaluation.}
\label{fig:scatter_plot_cluster_examples}
\end{figure*}

\begin{figure*}[t]
\centering
    \includegraphics[width=0.19\linewidth]{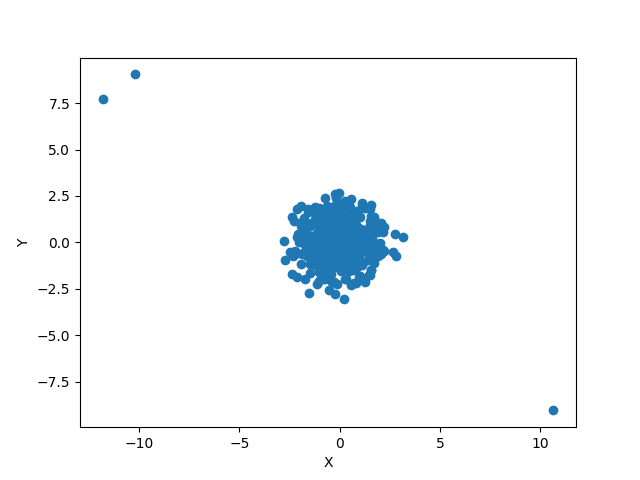}
    \includegraphics[width=0.19\linewidth]{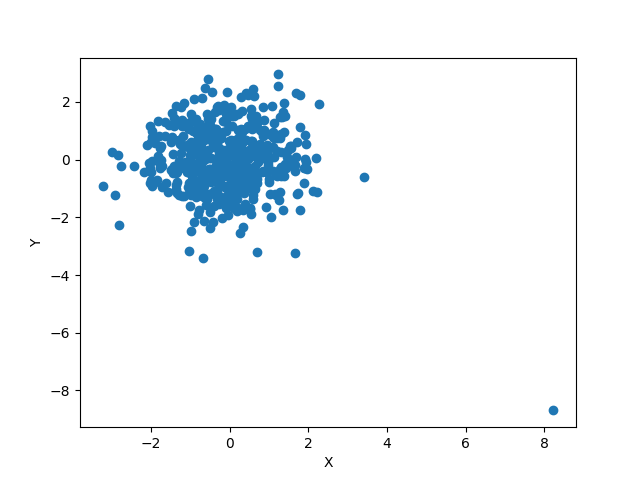}
    \includegraphics[width=0.19\linewidth]{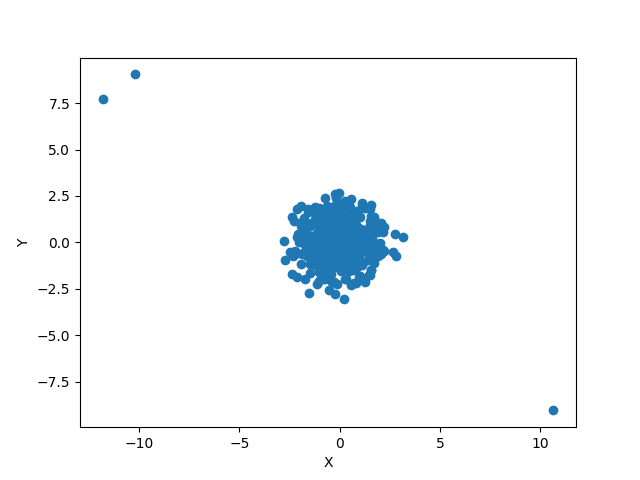}
    \includegraphics[width=0.19\linewidth]{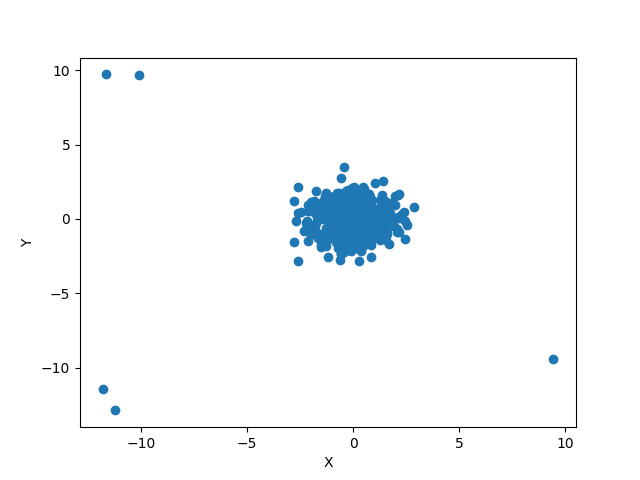}
    \includegraphics[width=0.19\linewidth]{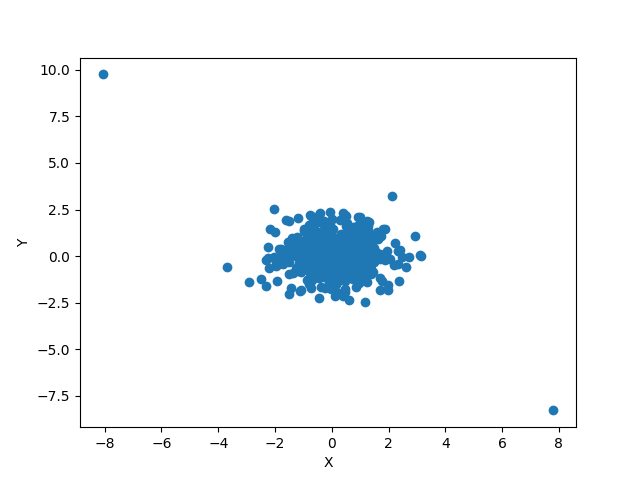}
    \includegraphics[width=0.19\linewidth]{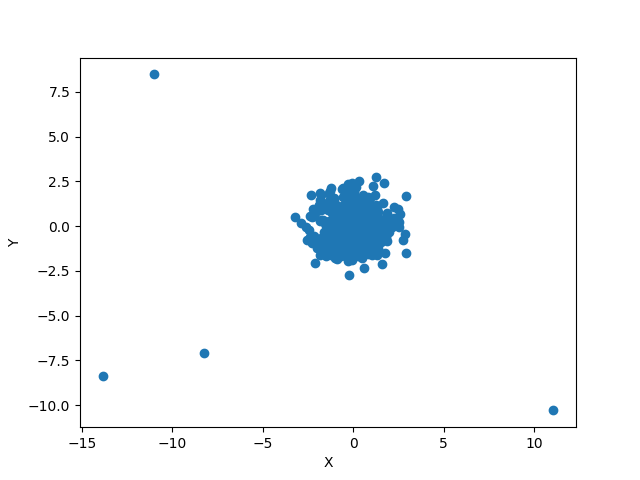}
    \includegraphics[width=0.19\linewidth]{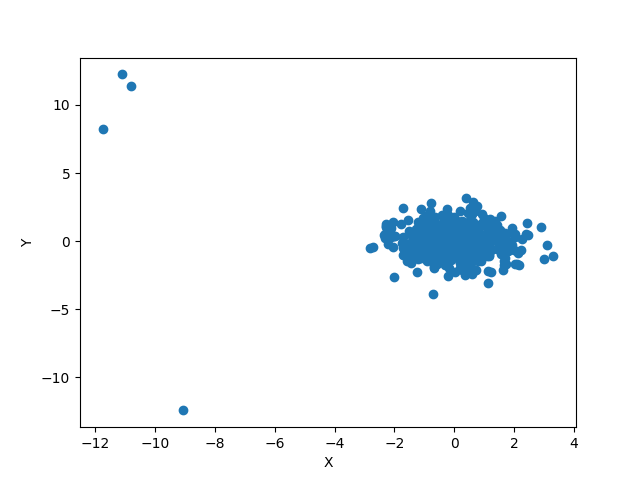}
    \includegraphics[width=0.19\linewidth]{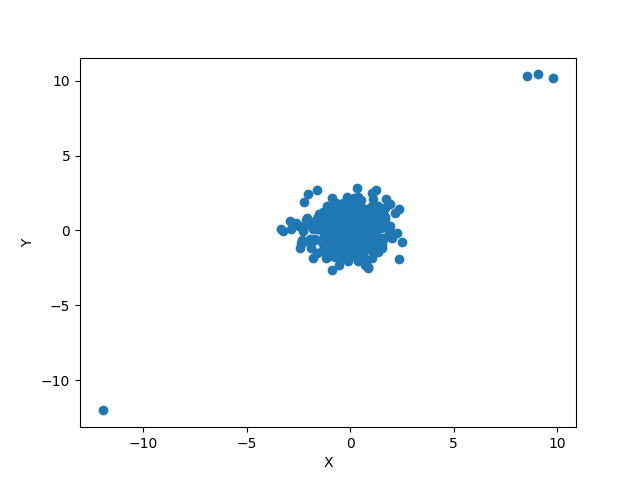}
    \includegraphics[width=0.19\linewidth]{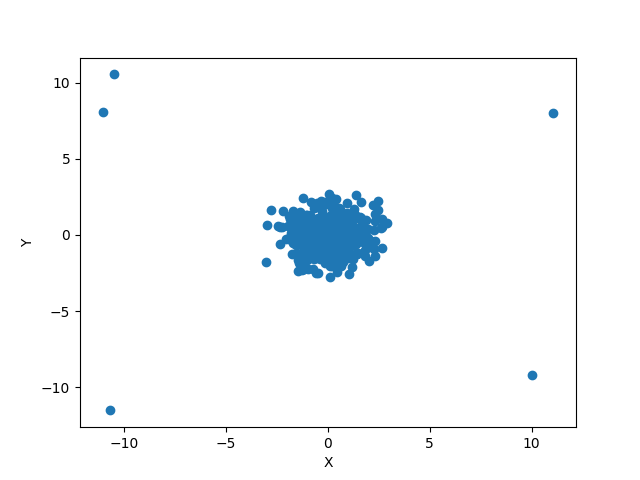}
    \includegraphics[width=0.19\linewidth]{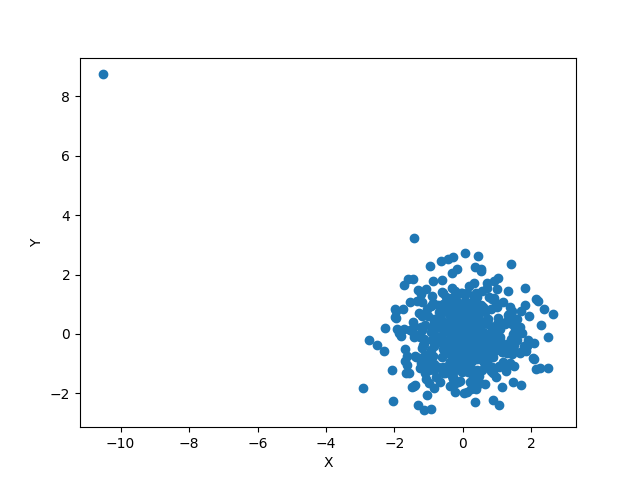}

    \caption{The randomly generated scatter plot with different numbers of outliers for LLM evaluation.}
\label{fig:scatter_plot_outlier_examples}
\end{figure*}

\begin{figure*}[t]
\centering
    \includegraphics[width=0.19\linewidth]{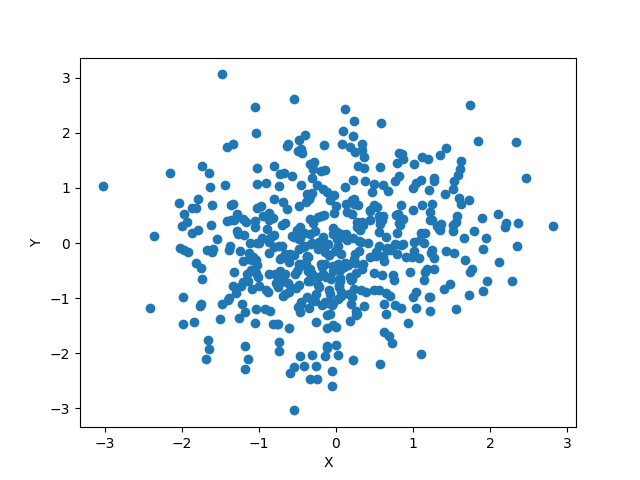}
    \includegraphics[width=0.19\linewidth]{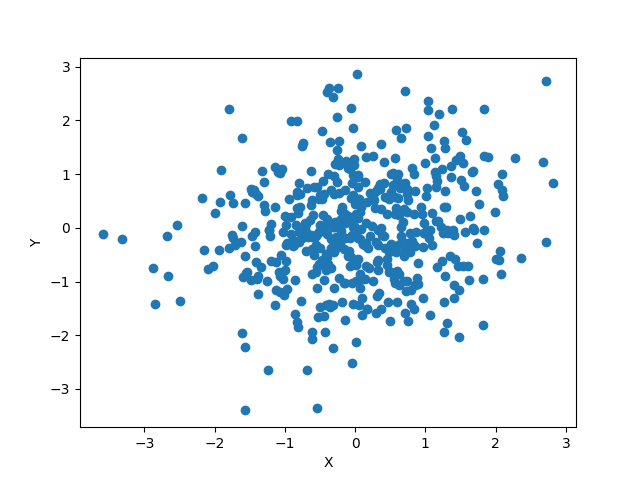}
    \includegraphics[width=0.19\linewidth]{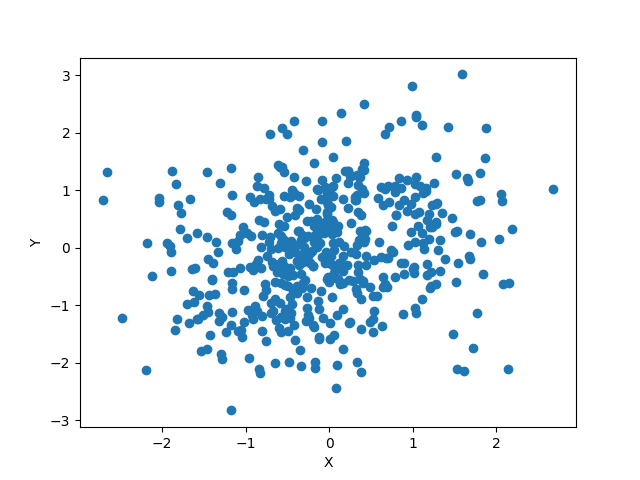}
    \includegraphics[width=0.19\linewidth]{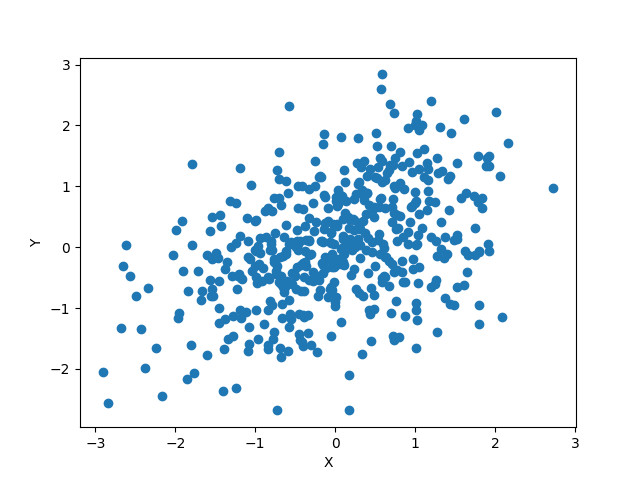}
    \includegraphics[width=0.19\linewidth]{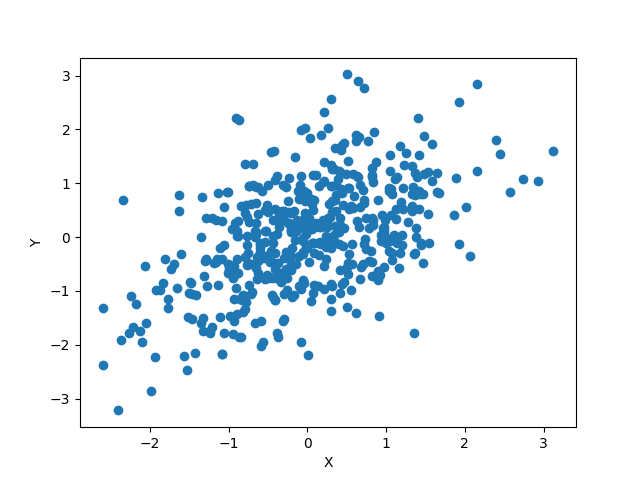}
    \includegraphics[width=0.19\linewidth]{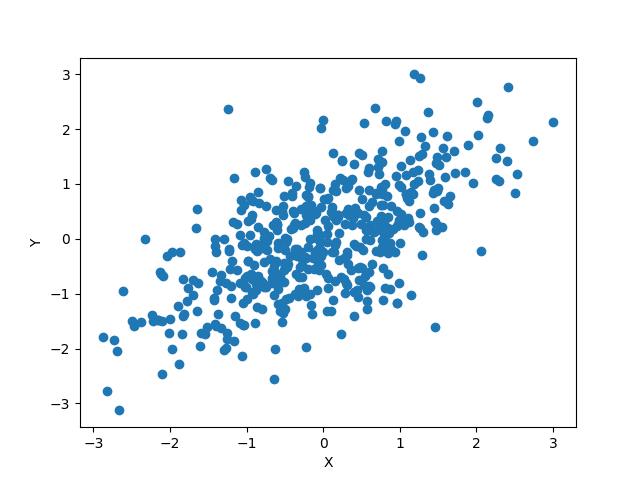}
    \includegraphics[width=0.19\linewidth]{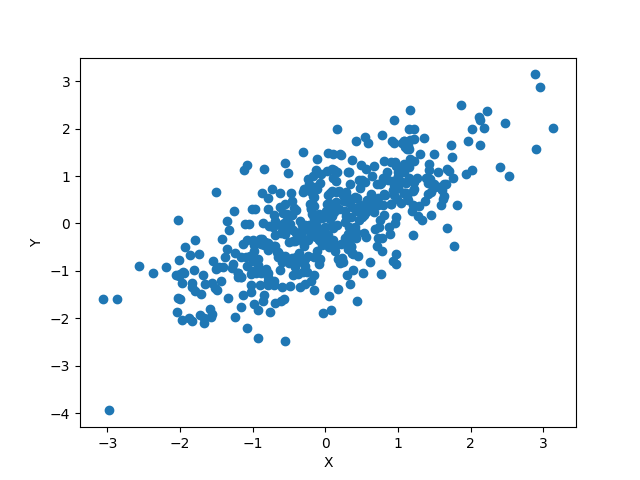}
    \includegraphics[width=0.19\linewidth]{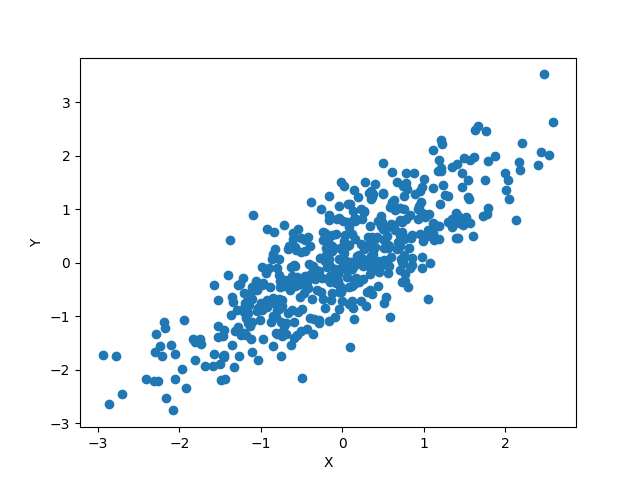}
    \includegraphics[width=0.19\linewidth]{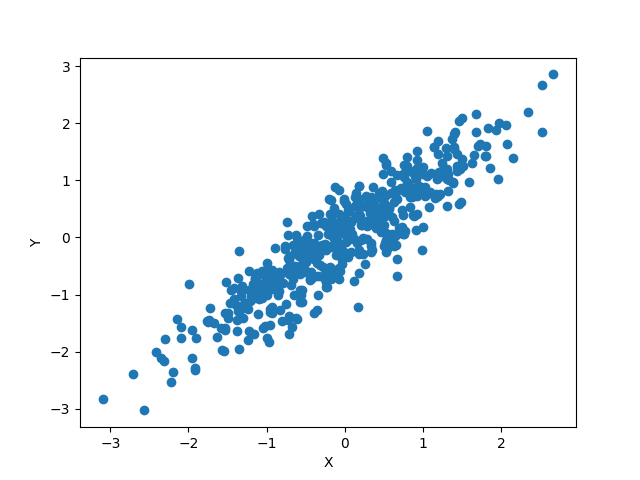}
    \includegraphics[width=0.19\linewidth]{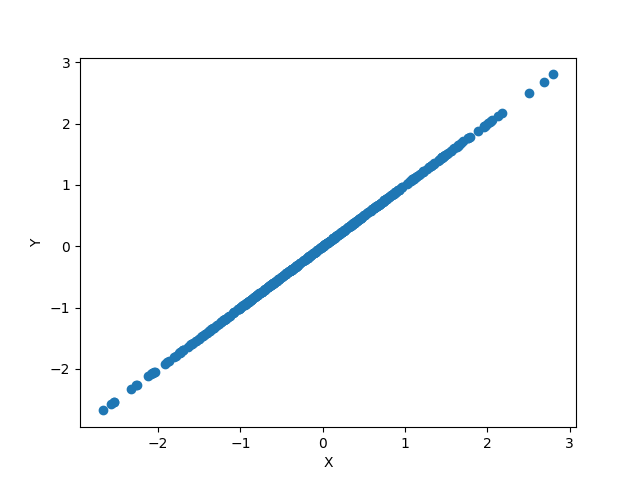}

    \caption{The randomly generated scatter plot with different correlation coefficients.}
\label{fig:scatter_plot_correlation_examples}
\end{figure*}

\subsection{Parallel Coordinate}
In parallel coordinate experiments, each task will be performed in 10 experiments, and the final results are aggregated as a percentage number (success rate)
In each experiment, 500 points are generated randomly following a pre-defined pattern (e.g., number of clusters).
\begin{itemize}
    \item \textbf{Clustering Count:}
    For the clustering task, we generate 1 to 10 clusters for each task.
    
    \textit{\textbf{Prompts}: "You are a parallel coordinate visualization expert. Is there any cluster in this visualization? Can you tell me how many clusters are in this visualization?"}

    \item \textbf{Outlier Count:} 
    Different from the cluster recognition task,  the outlier detection will sample 1-5 outlier points without overlap in the visualization.
    
    \textit{\textbf{Prompts}: "You are a parallel coordinate visualization expert. Is there any outlier in this visualization? Can you tell me how many outliers are in this visualization?"}

    \item \textbf{Correlation Detection} 
    
    Compared with the scatter plot visualization, in the correlation task, we randomly select two attributions to be correlated and these two attributions are nearby in the parallel coordinate visualization.

    \textit{\textbf{Prompts}: "You are a parallel coordinate visualization expert. Is there any correlation between these variables?}
\end{itemize}
\begin{figure*}[t]
\centering
    \includegraphics[width=0.19\linewidth]{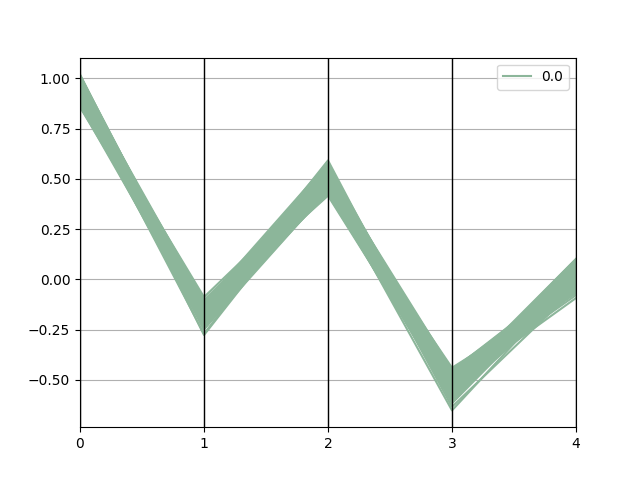}
    \includegraphics[width=0.19\linewidth]{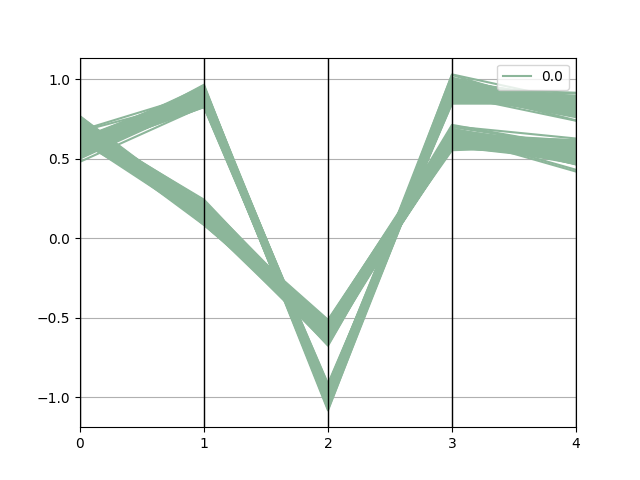}
    \includegraphics[width=0.19\linewidth]{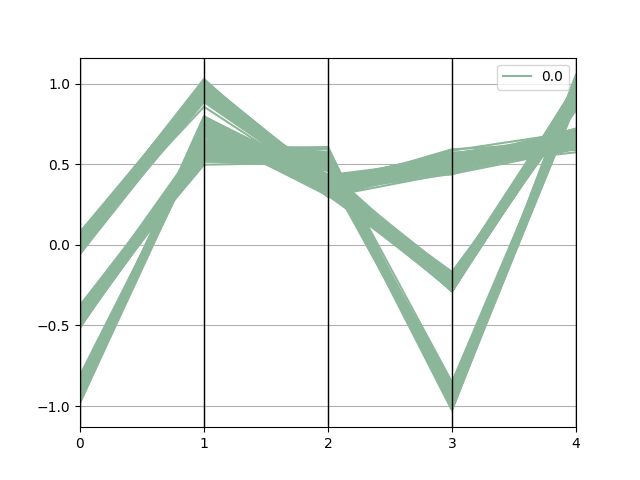}
    \includegraphics[width=0.19\linewidth]{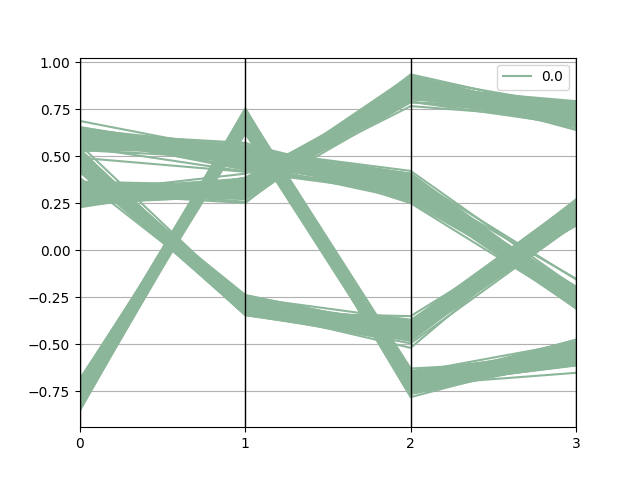}
    \includegraphics[width=0.19\linewidth]{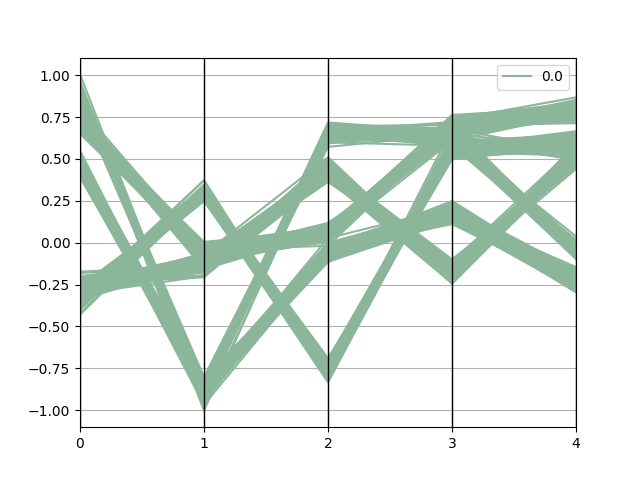}
    \includegraphics[width=0.19\linewidth]{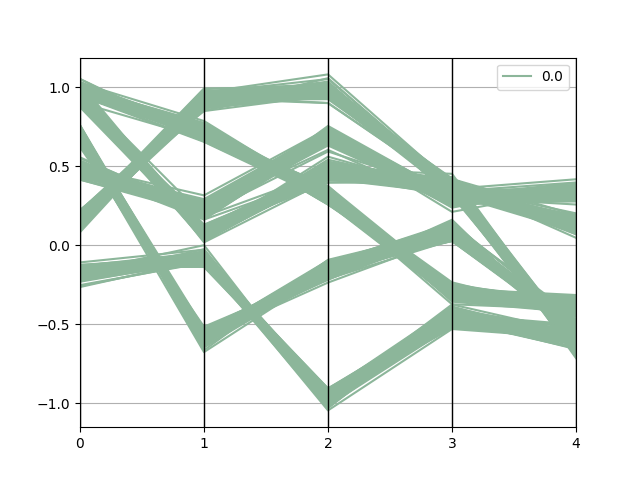}
    \includegraphics[width=0.19\linewidth]{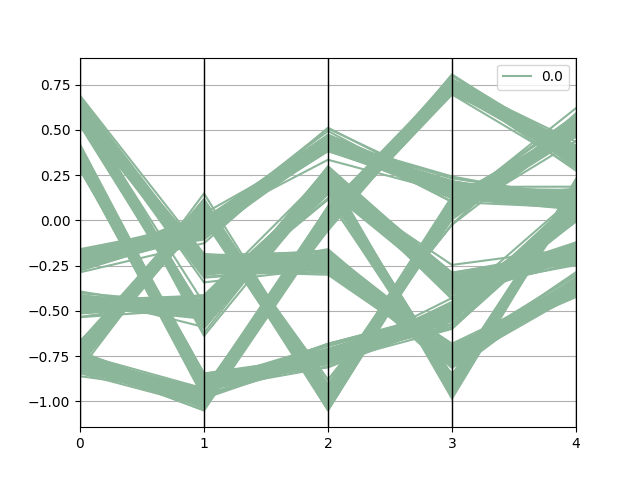}
    \includegraphics[width=0.19\linewidth]{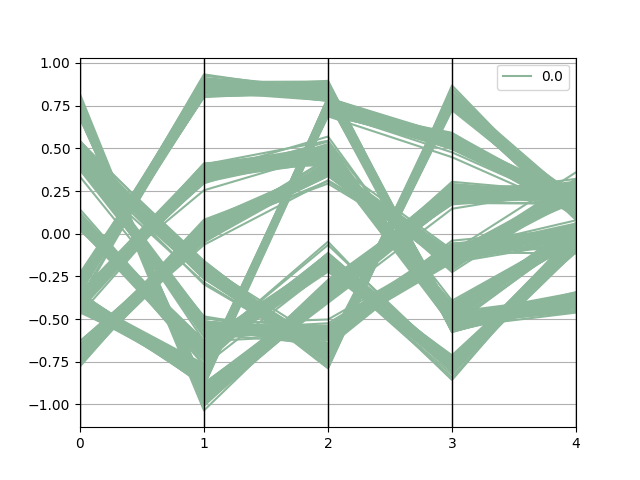}
    \includegraphics[width=0.19\linewidth]{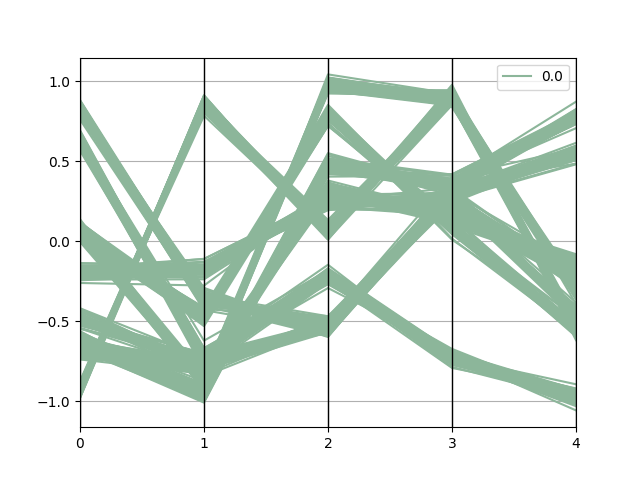}
    \includegraphics[width=0.19\linewidth]{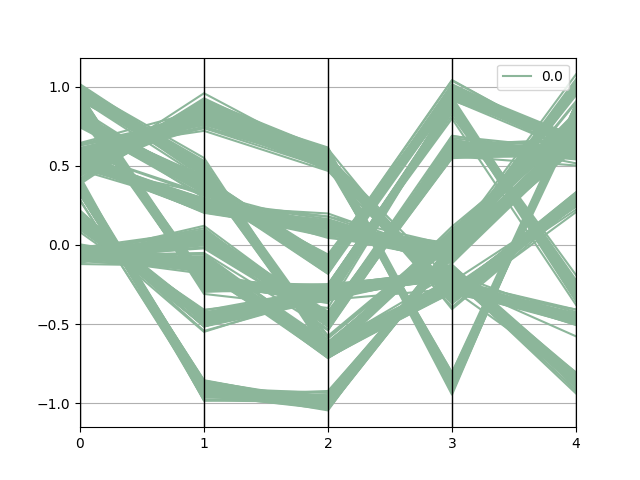}

    \caption{The randomly generated parallel coordinate with five dimensions and different numbers of clusters for LLM evaluation.}
\label{fig:parallel_coordinate_cluster_examples}
\end{figure*}
\begin{figure*}[t]
\centering
    \includegraphics[width=0.19\linewidth]{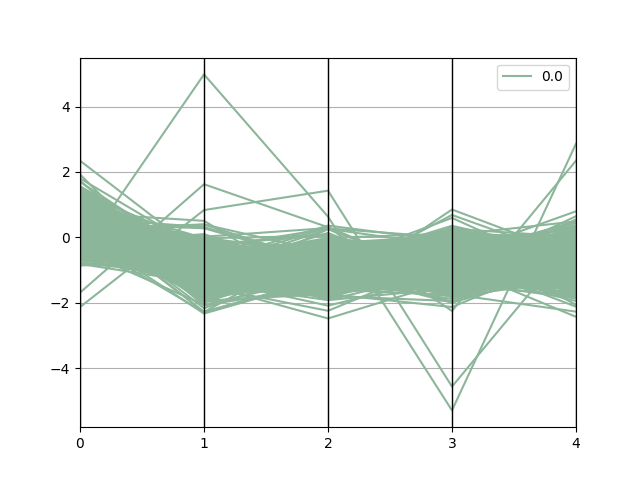}
    \includegraphics[width=0.19\linewidth]{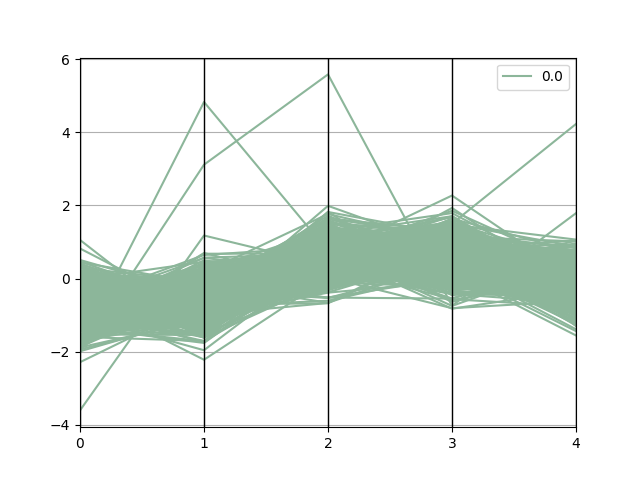}
    \includegraphics[width=0.19\linewidth]{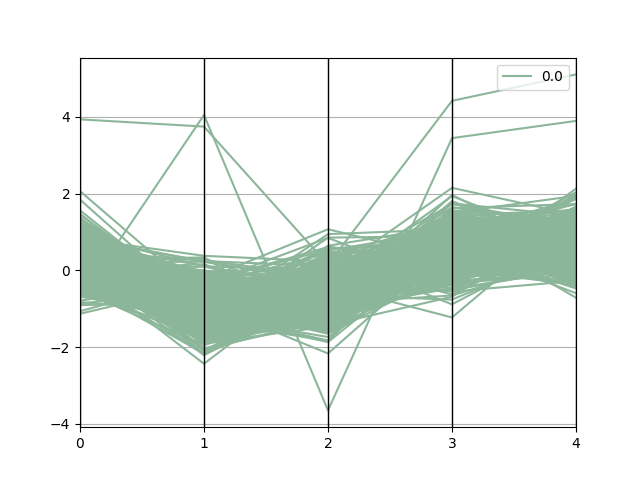}
    \includegraphics[width=0.19\linewidth]{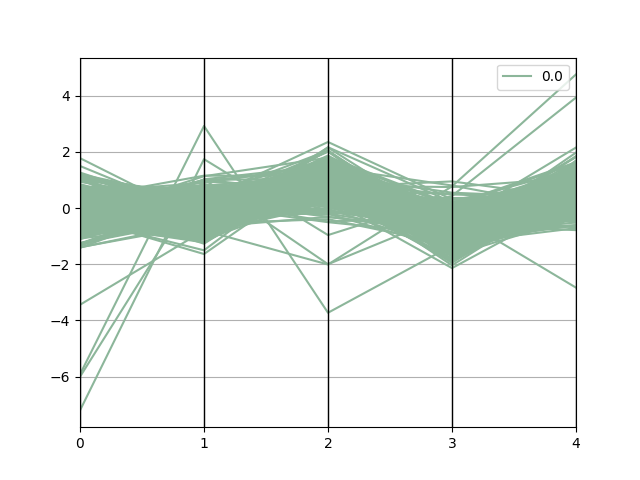}
    \includegraphics[width=0.19\linewidth]{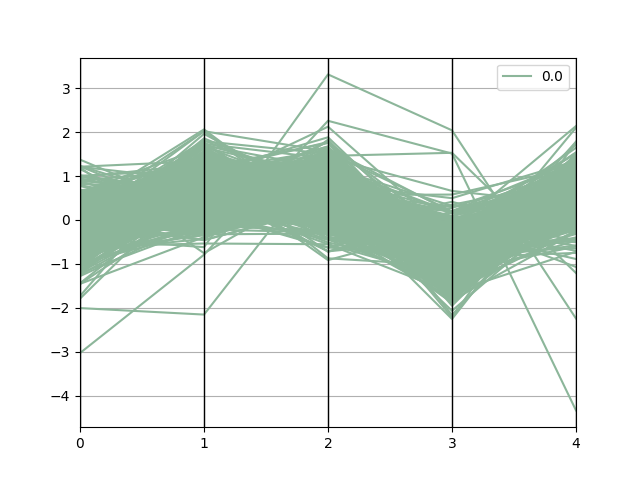}
    \includegraphics[width=0.19\linewidth]{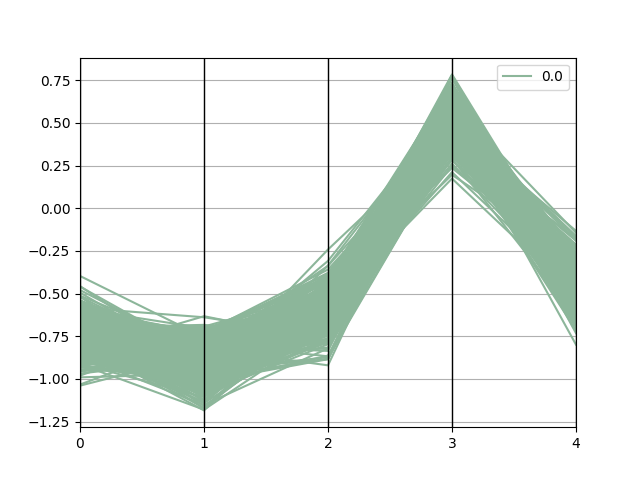}
    \includegraphics[width=0.19\linewidth]{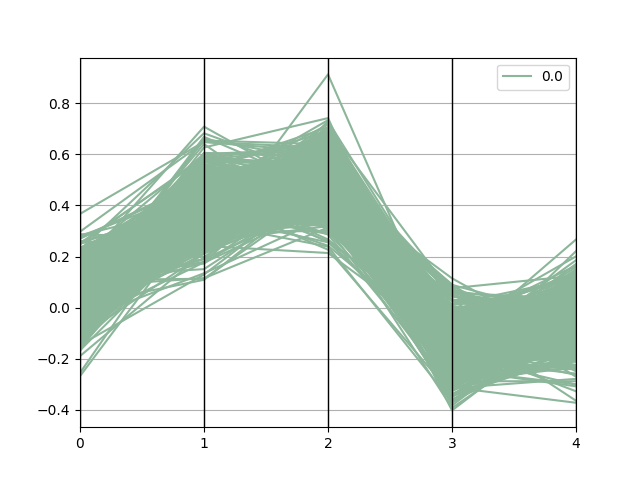}
    \includegraphics[width=0.19\linewidth]{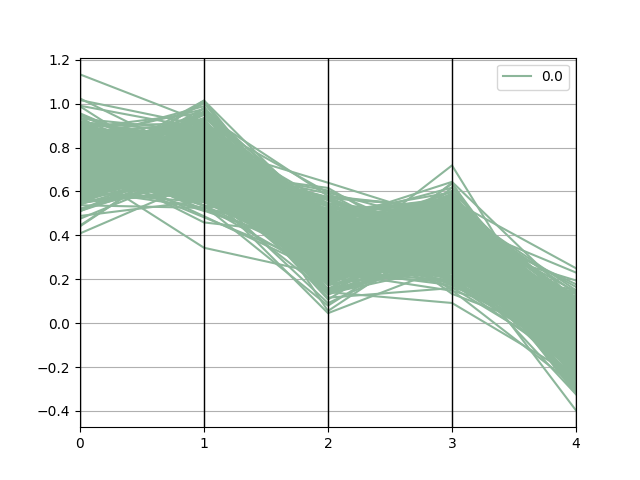}
    \includegraphics[width=0.19\linewidth]{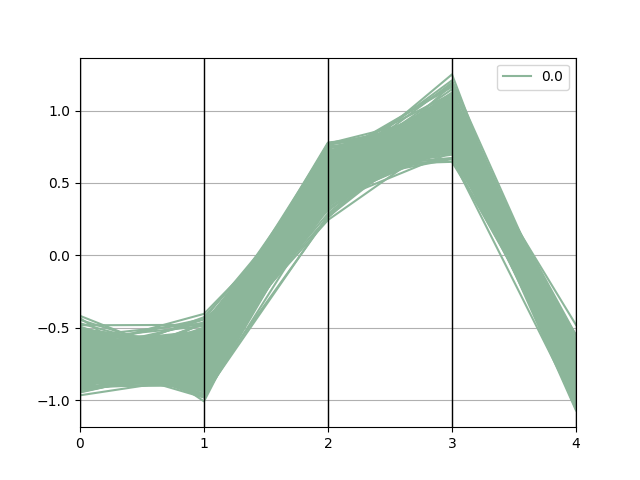}
    \includegraphics[width=0.19\linewidth]{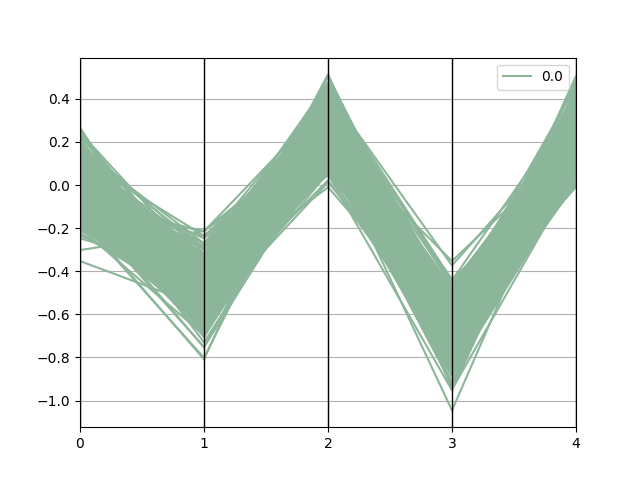}

    \caption{The randomly generated parallel coordinate with five dimensions and different numbers of outliers for LLM evaluation.}
\label{fig:parallel_coordinate_outlier_examples}
\end{figure*}
\begin{figure*}[t]
\centering
    \includegraphics[width=0.19\linewidth]{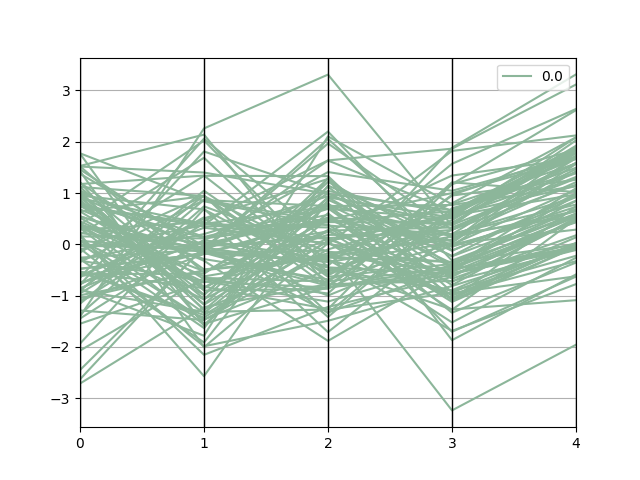}
    \includegraphics[width=0.19\linewidth]{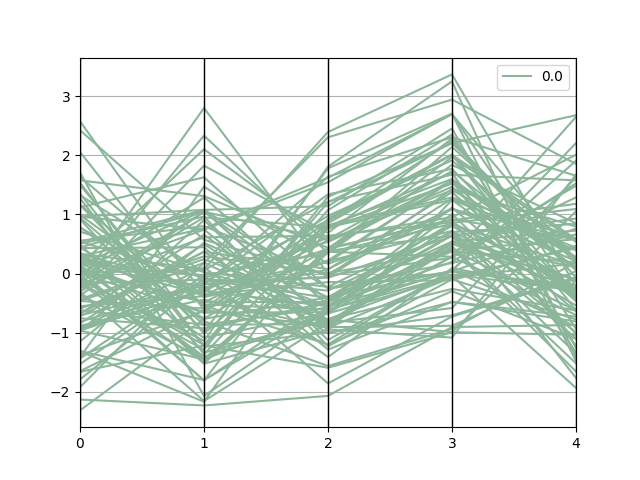}
    \includegraphics[width=0.19\linewidth]{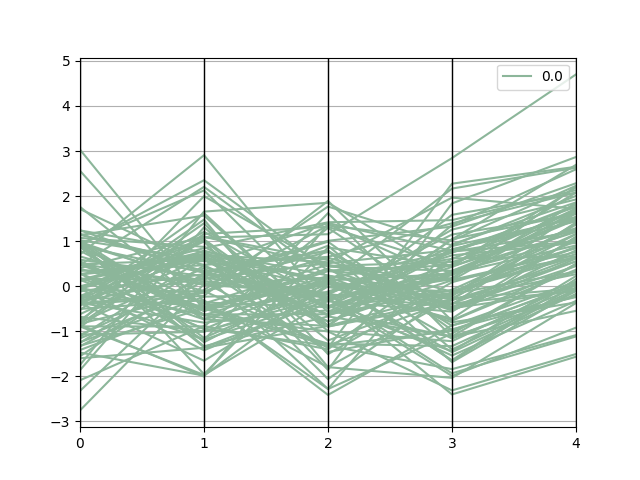}
    \includegraphics[width=0.19\linewidth]{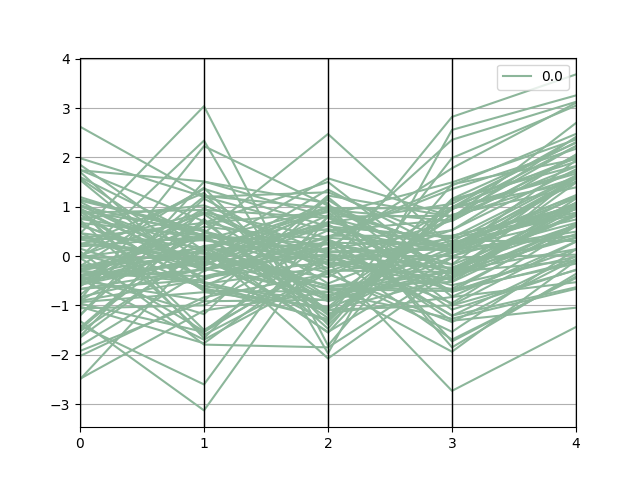}
    \includegraphics[width=0.19\linewidth]{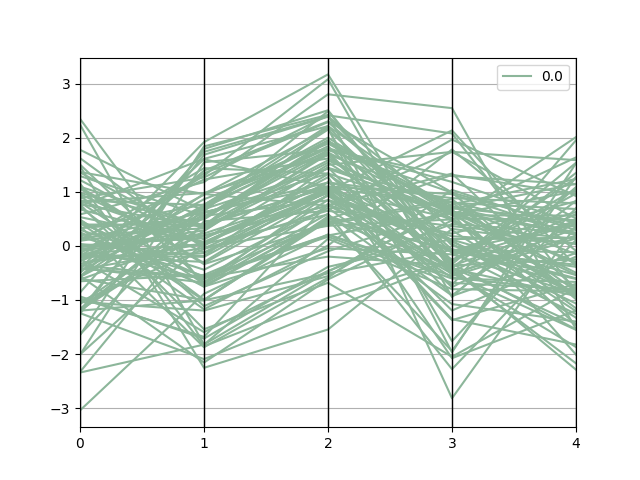}
    \includegraphics[width=0.19\linewidth]{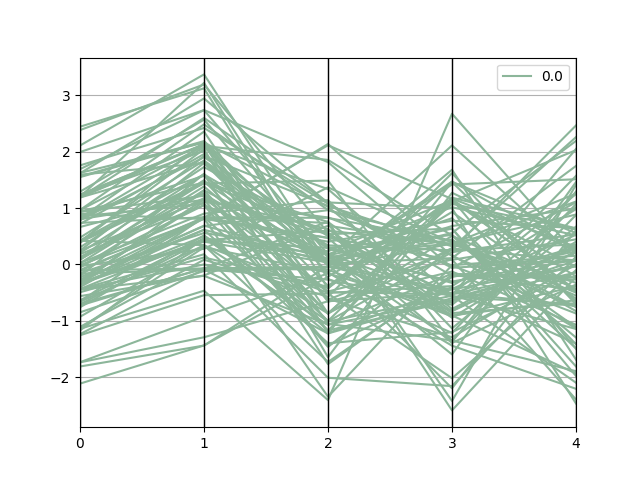}
    \includegraphics[width=0.19\linewidth]{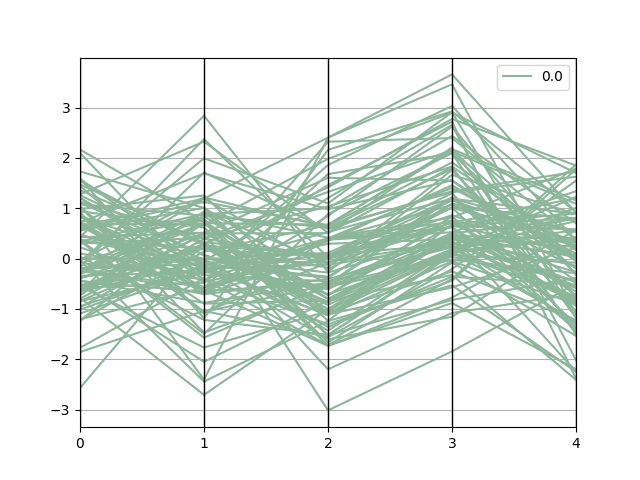}
    \includegraphics[width=0.19\linewidth]{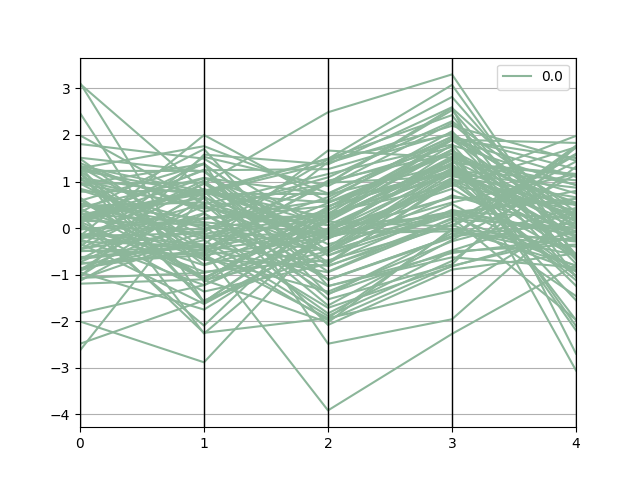}
    \includegraphics[width=0.19\linewidth]{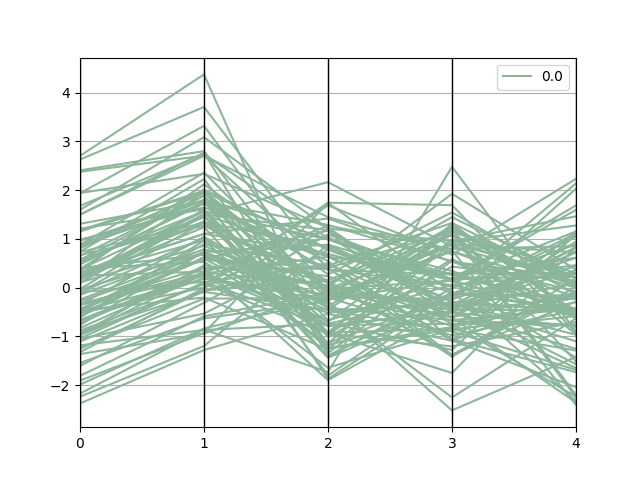}
    \includegraphics[width=0.19\linewidth]{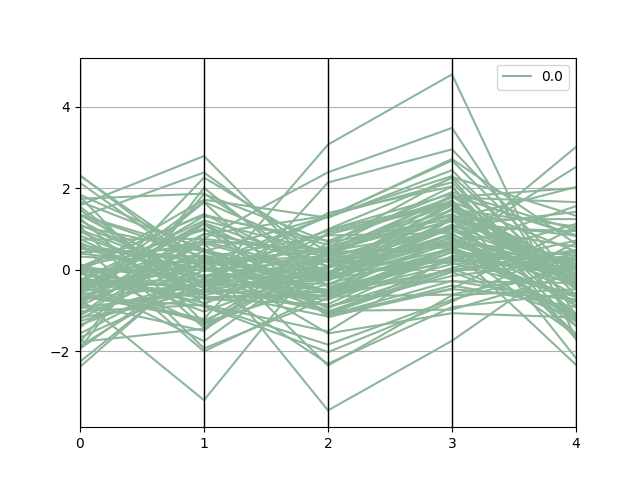}

    \caption{The randomly generated parallel coordinate with five dimensions and two of them have a high correlation.}
\label{fig:parallel_coordinate_correlation_examples}
\end{figure*}

\subsection{Graph}
It is worth noticing that we use the model to understand the result of an image without interaction operation.
Therefore, we only use graph visualization which is visually interpretable (e.g., not edge or node clutter).
We use a graph with 10 nodes and the overall sparsity is 20\%.
The final visualization is displayed by force-directed graph layout.
Similarly, each experiment will be performed 10 times and each time the graph and connection are randomly generated.

Prompts used for the graph in LLM evaluation:
\begin{itemize}
    \item \textbf{node count:} 
    \textit{Prompts: "You are a graph visualization expert. How many nodes are in this visualization?}
    
    \item \textbf{find node:} 
    \textit{Prompts: "You are a graph visualization expert. Is there a node named XXX in this visualization?}

    \item \textbf{connection:}
    \textit{Prompts:"You are a graph visualization expert. Is there a path from node XXX to node XXX?}

    \item \textbf{neighbor:}
    \textit{Prompts:"You are a graph visualization expert. What is the neighbor node of node XXX?}
\end{itemize}

\begin{figure*}[t]
\centering
    \includegraphics[width=0.19\linewidth]{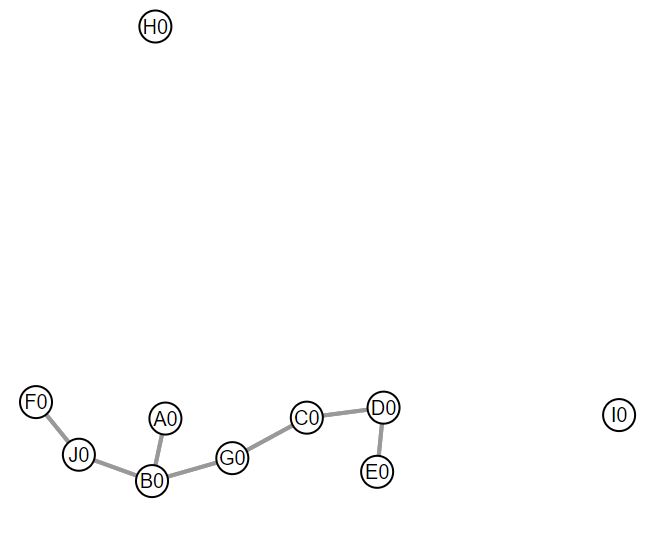}
    \includegraphics[width=0.19\linewidth]{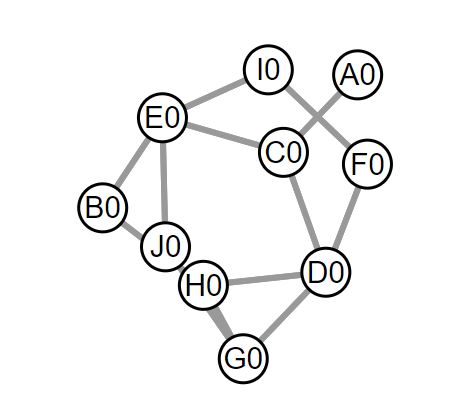}
    \includegraphics[width=0.19\linewidth]{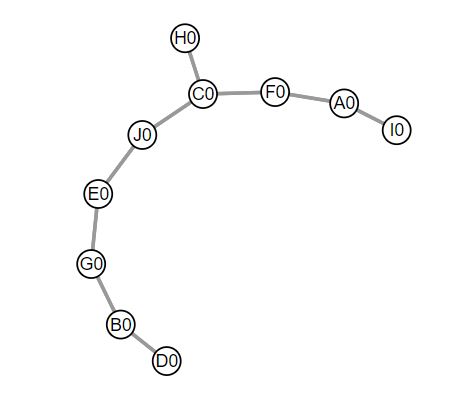}
    \includegraphics[width=0.19\linewidth]{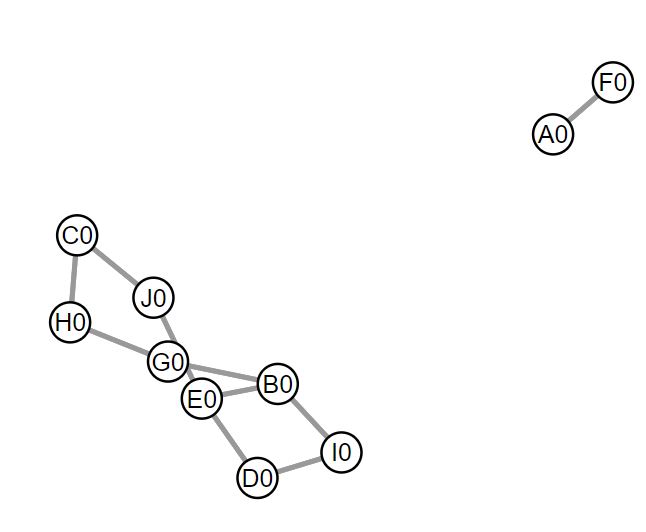}
    \includegraphics[width=0.19\linewidth]{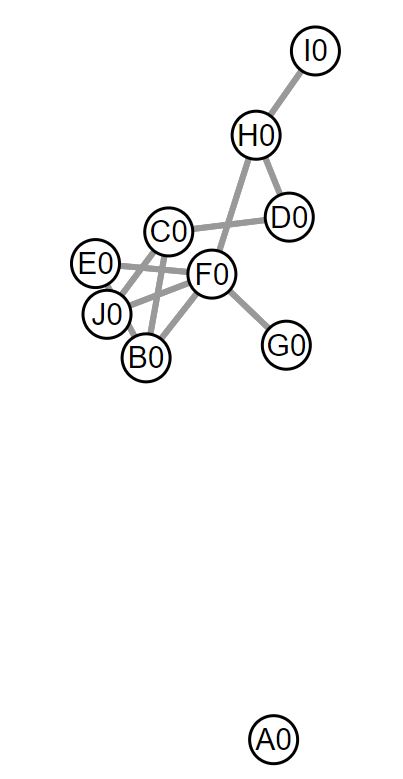}
    \includegraphics[width=0.19\linewidth]{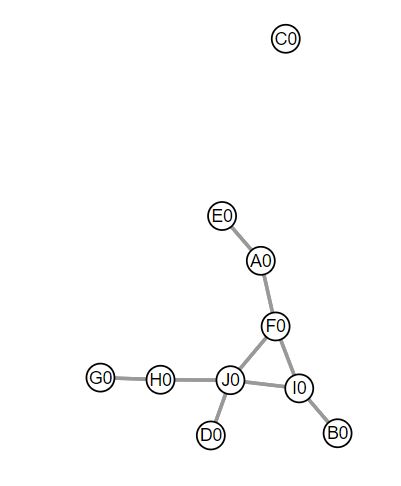}
    \includegraphics[width=0.19\linewidth]{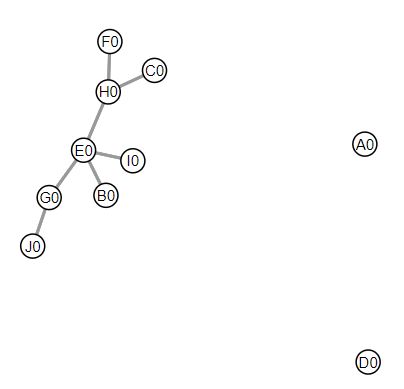}
    \includegraphics[width=0.19\linewidth]{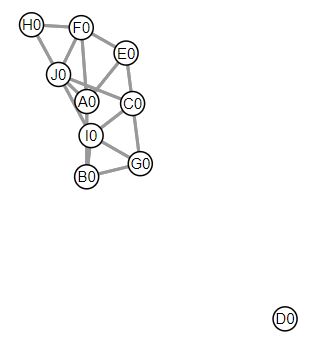}
    \includegraphics[width=0.19\linewidth]{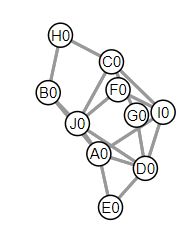}
    \includegraphics[width=0.19\linewidth]{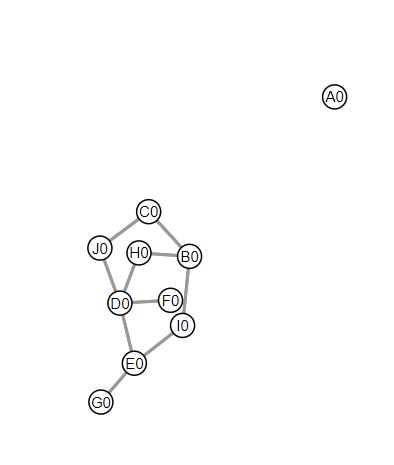}

    \caption{The randomly generated graph for graph exploration tasks}
\label{fig:graph_exploration}
\end{figure*}

\subsection{Volume Rendering}
Prompts used to evaluate the structure of interest recognition in the volume rendering: 

\begin{itemize}
    \item \textbf{Boston Teapot} 
    \textit{"You are provided with several screenshots showing a volume rendering of the same CT data, for each image assess whether you can recognize the structure of interest, a teapot. Only assess for the structure of interest and not any other structures you can recognize in the screenshot. Use only one of these options for assessment: 'Not recognizable', and 'Recognizable'. 'Not recognizable' means that the structure of interest cannot be identified in the image, even if another structure is recognizable.  'Recognizable' implies that both the structure of interest and its shape can be discerned in the screenshot.?}
    
    \item \textbf{Visible Male} 
    \textit{"You are provided with several screenshots showing a volume rendering of the same CT data, for each image assess whether you can recognize the structure of interest, a human face. Only assess for the structure of interest and not any other structures you can recognize in the screenshot. Use only one of these options for assessment: 'Not recognizable', and 'Recognizable'. 'Not recognizable' means that the structure of interest cannot be identified in the image, even if another structure is recognizable.  'Recognizable' implies that both the structure of interest and its shape can be discerned in the screenshot."}

\end{itemize}

\begin{figure}
  \centering
  \begin{subfigure}[t]{0.3\textwidth}
    \centering
    \includegraphics[width=\textwidth]{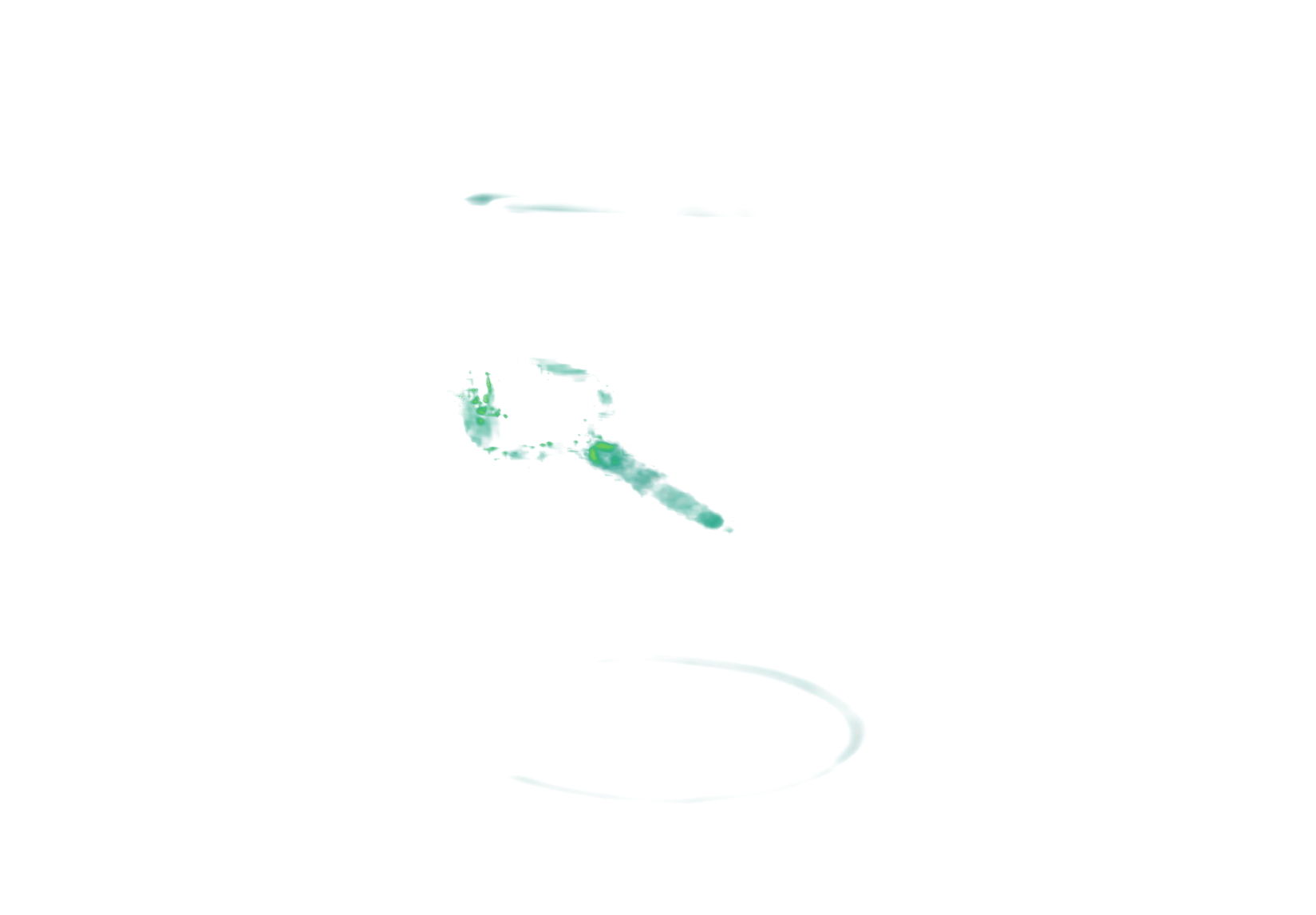}
    \caption{90\% }
    \label{fig:evalteapota}
  \end{subfigure}
  
  \begin{subfigure}{0.3\textwidth}
    \centering
    \includegraphics[width=\textwidth]{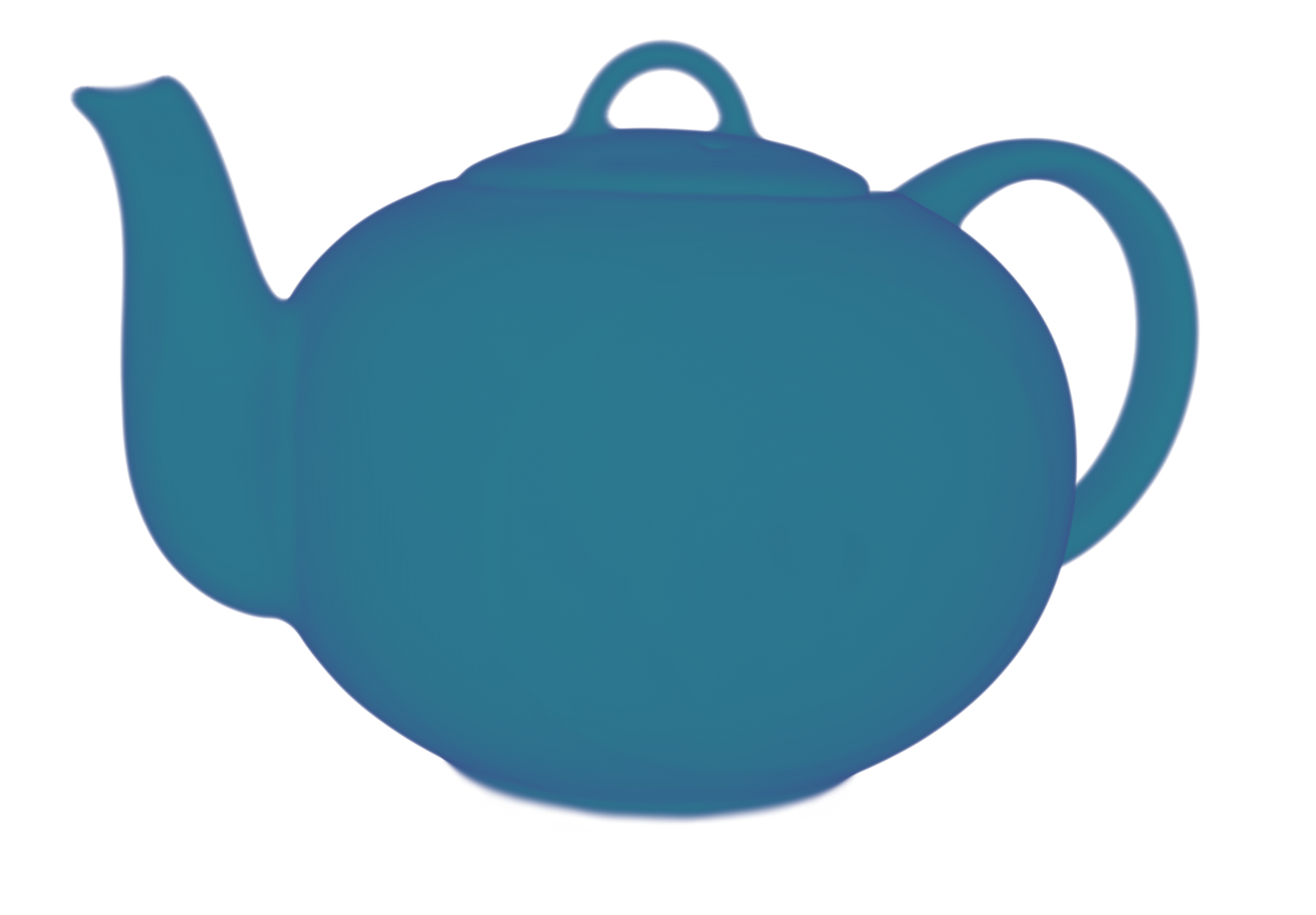}
    \caption{40\% }
    \label{fig:evalteapotb}
  \end{subfigure}

  \begin{subfigure}{0.3\textwidth}
    \centering
    \includegraphics[width=\textwidth]{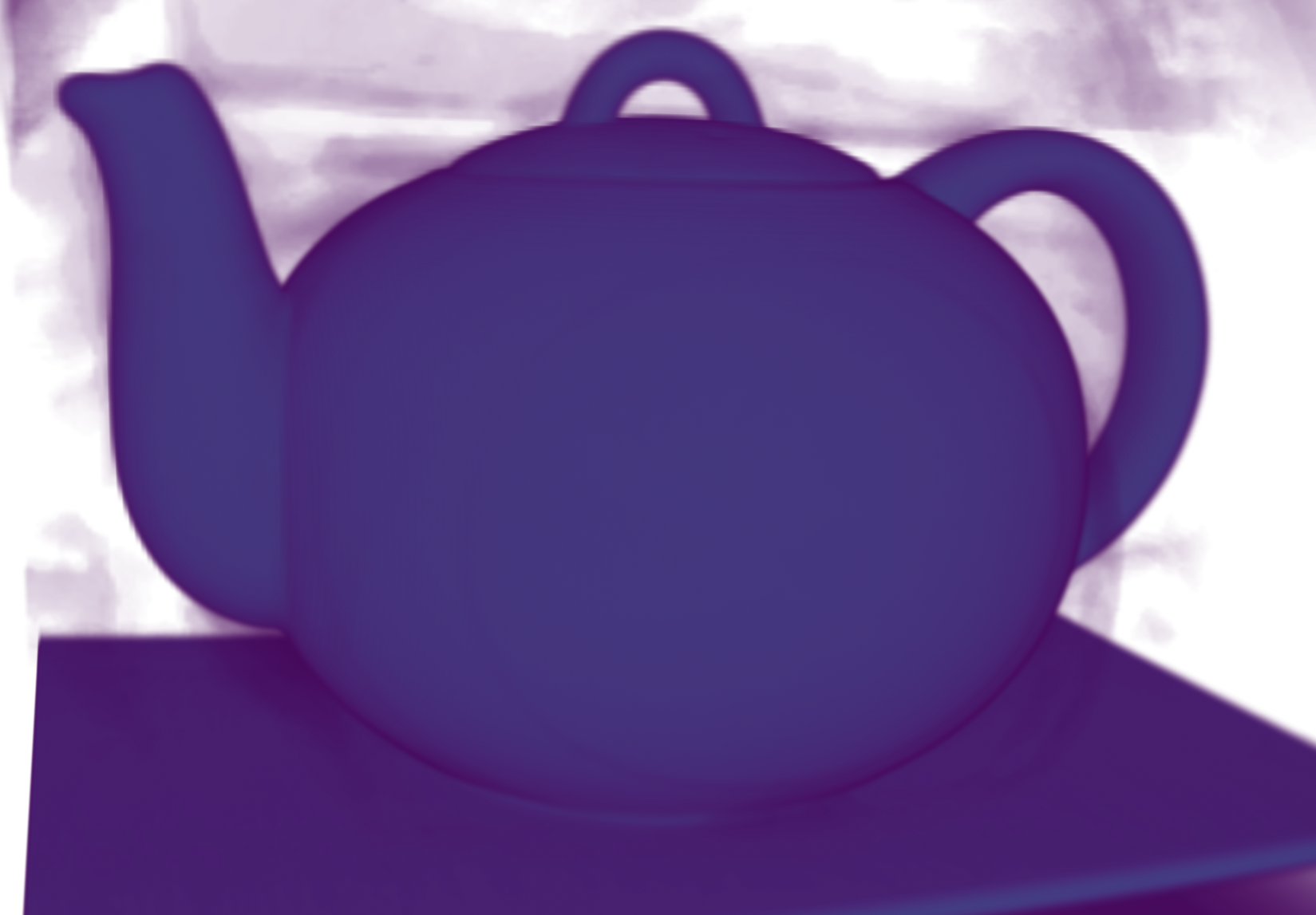}
    \caption{20\% }
    \label{fig:evalteapotc}
  \end{subfigure}

  \begin{subfigure}{0.3\textwidth}
    \centering
    \includegraphics[width=\textwidth]{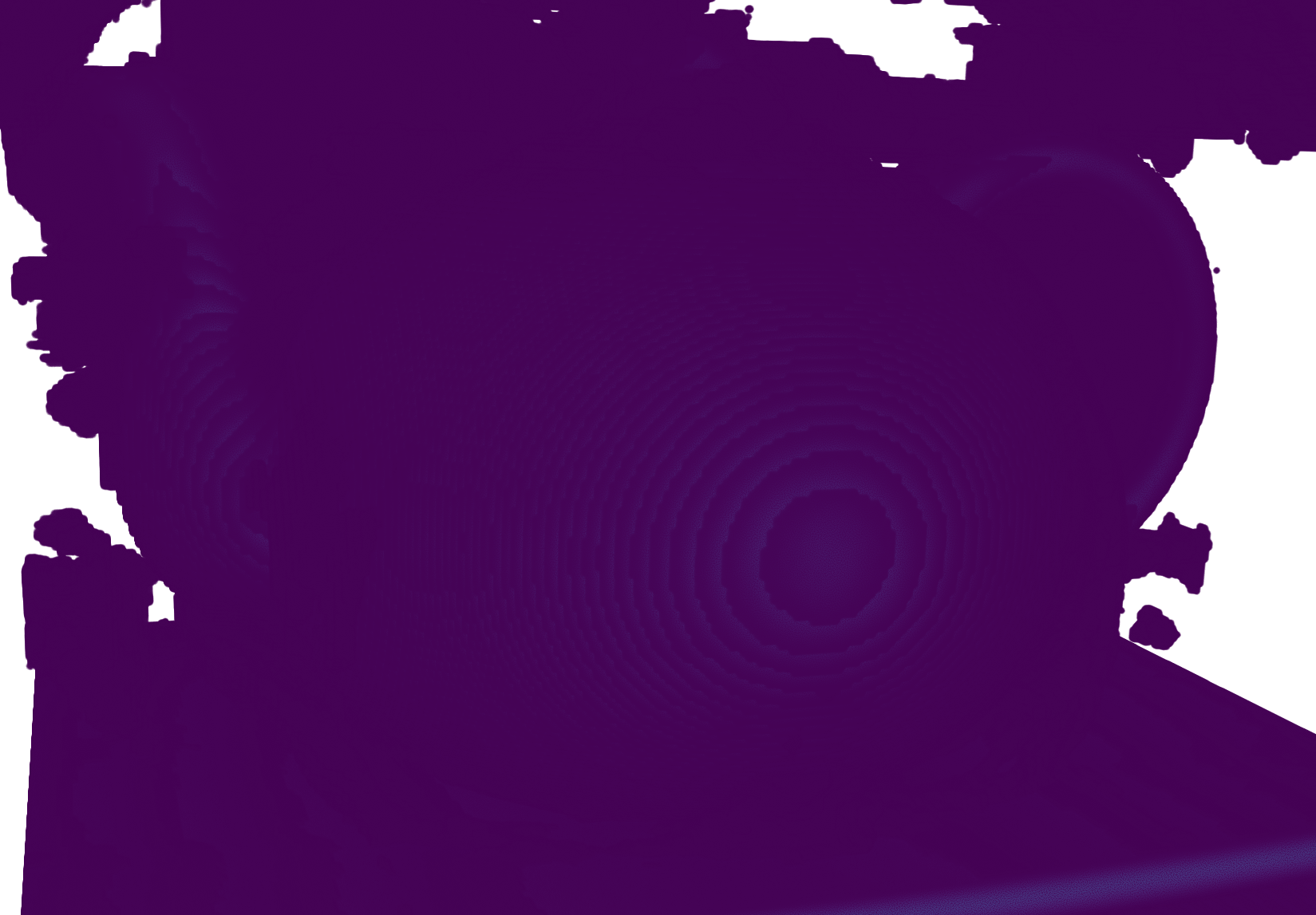}
    \caption{5\% }
    \label{fig:evalteapotd}
  \end{subfigure}
  
  \caption{The Boston Teapot dataset volume rendered using the same color map but at varying opacity levels. Structure of interest: the teapot. The response from the LLM model was \ref{fig:evalteapota}: 'not recognizable', \ref{fig:evalteapotb}: 'recognizable', \ref{fig:evalteapotc}: 'recognizable', and \ref{fig:evalteapotd}: 'not recognizable'}
  \label{fig:evalteapot}
\end{figure}

\begin{figure}[t]
  \centering
    \begin{subfigure}{0.23\textwidth}
    \centering
    \includegraphics[width=\textwidth]{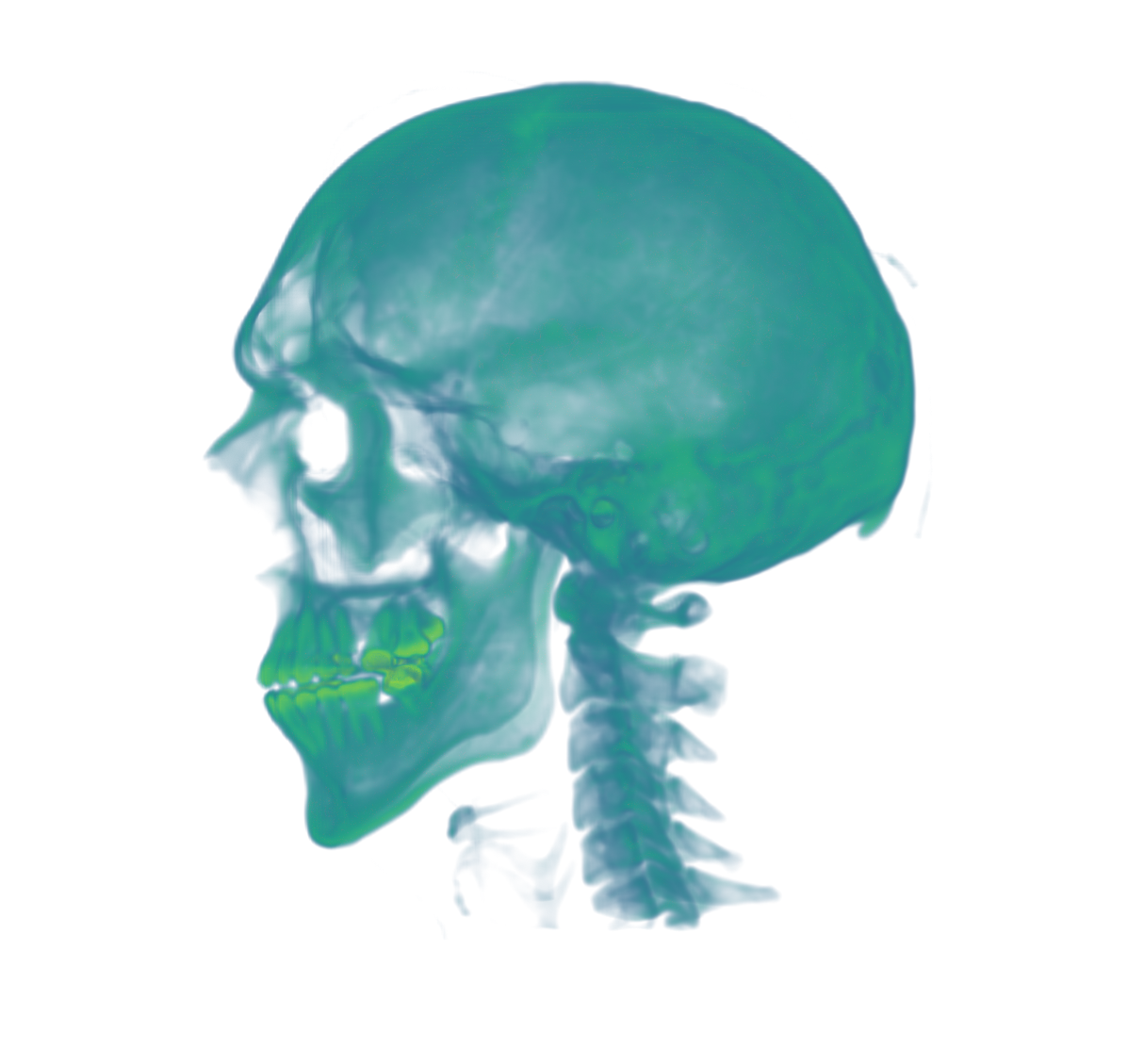}
    \subcaption{90\%}
    \label{fig:fig:evalhumanmalea}
    \end{subfigure}

    \begin{subfigure}{0.23\textwidth}
    \centering
    \includegraphics[width=\textwidth]{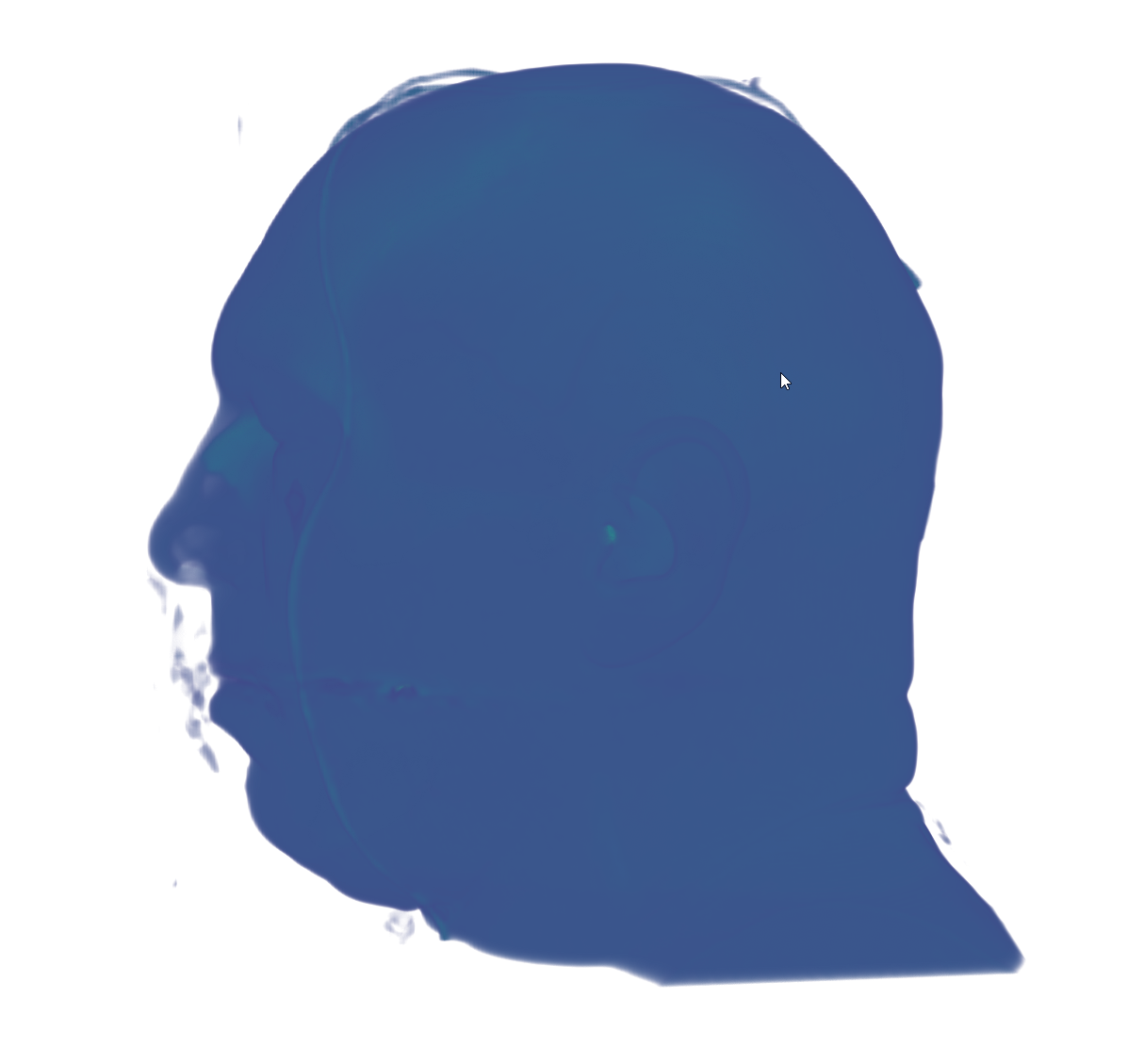}
    \caption{20\% }    \label{fig:evalhumanmaleb}
    \end{subfigure}

    \begin{subfigure}{0.23\textwidth}
    \centering
    \includegraphics[width=\textwidth]{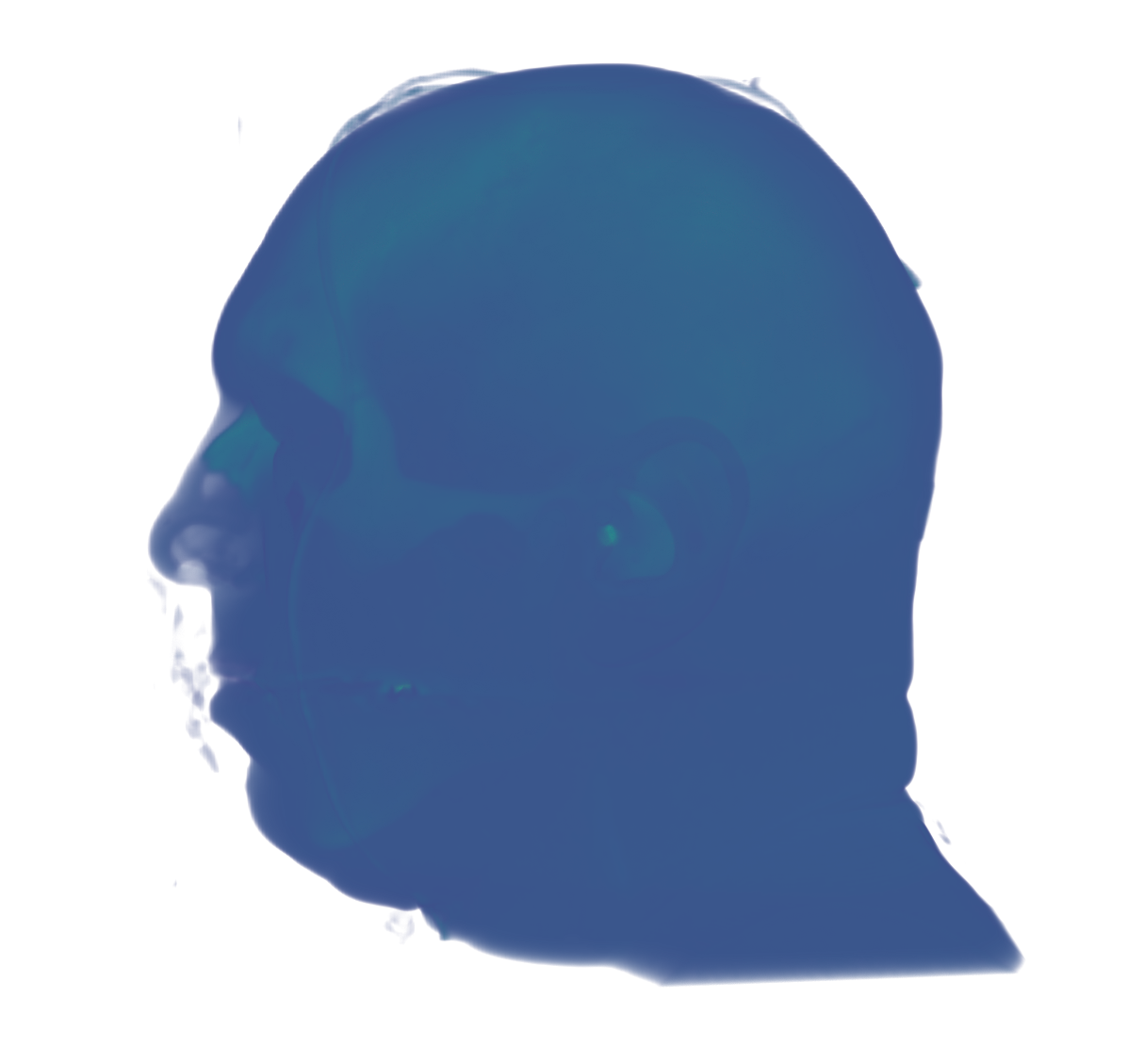}
    \caption{30\% }
    \label{fig:evalhumanmalec}
    \end{subfigure}
    
    \begin{subfigure}{0.23\textwidth}
    \centering
    \includegraphics[width=\textwidth]{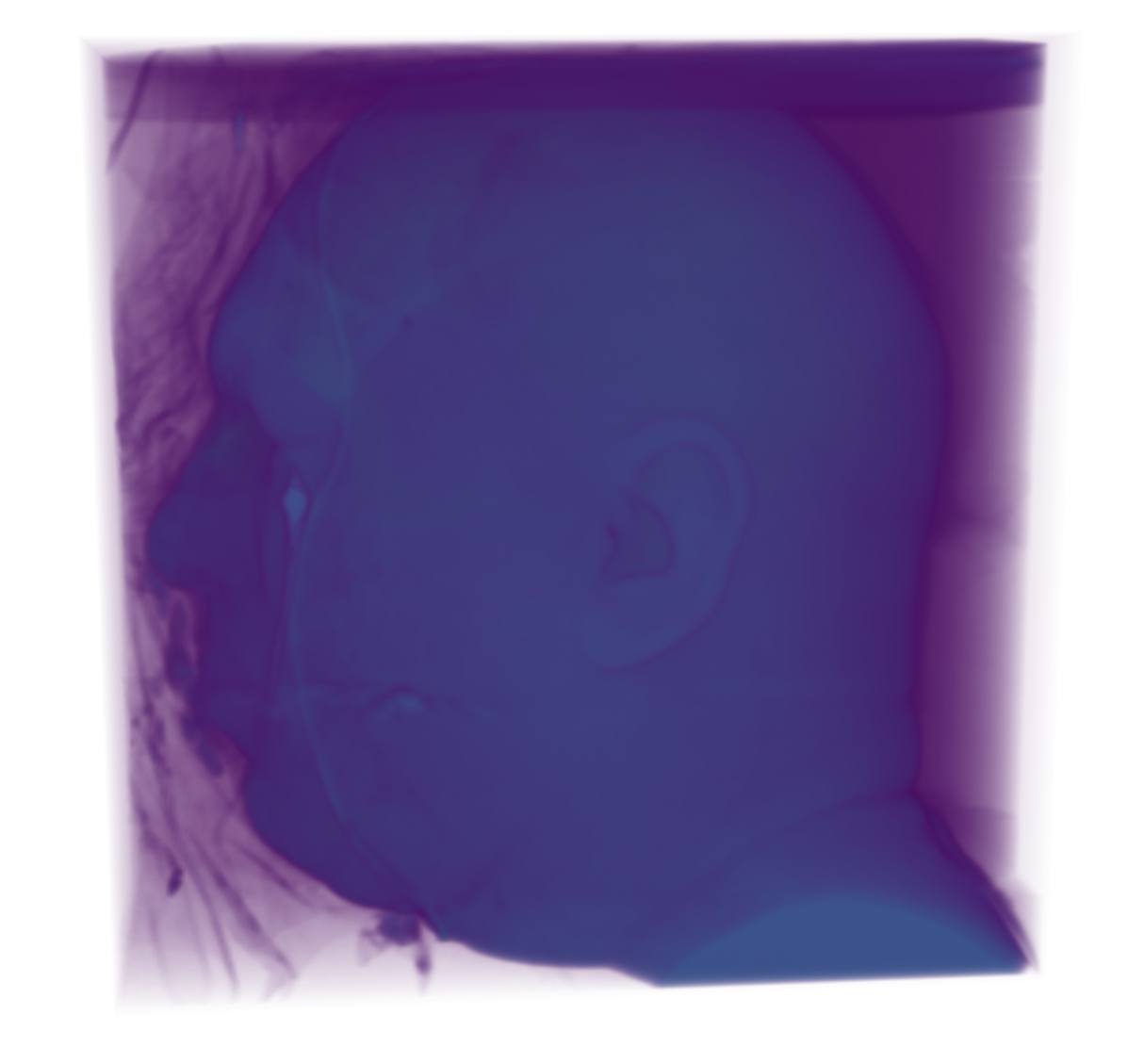}
    \caption{5\% }
    \label{fig:evalhumanmaled}
    \end{subfigure}
  
  \caption{The Human Male dataset.   In this figure, we conducted tests on the same set of screenshots, focusing on two distinct structures of interest: the male face and the bones. Notably, all images were accurately identified, except for the instance depicted in Figure \ref{fig:evalhumanmaled}, where the presence of background partially occludes the head. The high degree of noise in this scenario appears to have affected the recognition of the skull. The agent's response for a 'male face' as the structure of interest: \ref{fig:fig:evalhumanmalea}: 'not recognizable', \ref{fig:evalhumanmaleb}: 'recognizable', \ref{fig:evalhumanmalec}: 'recognizable', and \ref{fig:evalhumanmaled}: 'not recognizable'. The agent's response for a 'bones' structure of interest: \ref{fig:fig:evalhumanmalea}: 'recognizable', \ref{fig:evalhumanmaleb}: 'not recognizable', \ref{fig:evalhumanmalec}: 'not recognizable', and \ref{fig:evalhumanmaled}: 'recognizable'. }
  \label{fig:evalhumanmale}
\end{figure}


\begin{algorithm}
  \caption{Opacity Transfer Function Adjustment - Used for the Heuristic-centric Action Plan}
      \begin{algorithmic}[1]
      
      \State $scalar\_range \gets max\_val - min\_val$
      \State $window\_width \gets scalar\_range / bins$
      \State $step\_size \gets window\_width \times window\_factor$
      \State $start\_point \gets min\_val$
      \State $end\_point \gets start\_point + window\_width$
      \Repeat
        \State $assessment\_result \gets assess\_screenshot()$
        \If {$assessment\_result$ == "not recognizable"}
          \State $start\_point \gets start\_point + step\_size$
          \State $end\_point \gets end\_point + step\_size$
        \ElsIf {$assessment\_result$ == "recognizable"}
          \State $start\_point \gets start\_point + step\_size \times speed\_reduction$
          \State $end\_point \gets end\_point + step\_size \times speed\_reduction$
        \EndIf
      \Until {$assessment\_result$ == "clear"}
      \end{algorithmic}
      
  \end{algorithm}

\end{document}